\documentclass[12pt,a4paper]{article}
\usepackage{graphicx,amssymb,amsmath,color,scalefnt}
\usepackage{placeins}
\usepackage{multirow, makecell}
\usepackage{soul}
\usepackage{chngpage}
\usepackage{geometry}
\usepackage[dvipsnames]{xcolor}
\usepackage{rotating}
\usepackage[normalem]{ulem}
\usepackage{jheppub}
\usepackage{scalerel}
\usepackage{makecell}
\usepackage{appendix}
\renewcommand{\thefootnote}{\fnsymbol{footnote}}

\makeatletter\g@addto@macro\bfseries{\boldmath}\makeatother

\DeclareMathAlphabet{\mathsfit}{\encodingdefault}{\sfdefault}{m}{sl}

\newcommand{\FDF}[1][]{\varphi^\dagger #1\!\overleftrightarrow{D}\!_\mu\varphi}
\newcommand{\FDFI}[1][]{\varphi^\dagger #1\!\overleftrightarrow{D}^I\!\!\!_\mu\:\varphi}
\newcommand{\mailto}[1]{\href{mailto:#1}{#1}}

\begin{document}

\title{Snowmass White Paper: prospects for the measurement of top-quark couplings}

\author[a]{Gauthier Durieux,}
\author[b]{Abel Guti\'errez Camacho,}
\author[c]{Luca Mantani,}
\author[b,d]{V\'ictor Miralles\footnote{Corresponding author: \mailto{victor.miralles@roma1.infn.it}},}%
\author[b]{Marcos Miralles L\'opez,}%
\author[b]{Mar\'ia Moreno Ll\'acer,}%
\author[e]{Ren\'e Poncelet,}
\author[f]{Eleni Vryonidou,}
\author[b]{Marcel Vos}

\affiliation[a]{CERN, Theoretical Physics Department, Geneva 23 CH-1211, Switzerland}
\affiliation[b]{IFIC, Universitat de Val\`encia and CSIC, Spain}
\affiliation[c]{DAMTP, University of Cambridge, Wilberforce Road, Cambridge, CB3 0WA, United Kingdom}
\affiliation[d]{INFN, Sezione di Roma, Piazzale A. Moro 2, I-00185 Roma, Italy}
\affiliation[e]{Cavendish Laboratory, University of Cambridge, Cambridge, CB3 0HE, United Kingdom}
\affiliation[f]{Department of Physics and Astronomy, University of Manchester, Oxford Road, Manchester M13 9PL, United Kingdom}

\def\TeV{\ifmmode {\mathrm{Te\kern -0.1em V}}\else
                   \textrm{Te\kern -0.1em V}\fi\,}%
\def\GeV{\ifmmode {\mathrm{Ge\kern -0.1em V}}\else
                   \textrm{Ge\kern -0.1em V}\fi\,}%
\def\MeV{\ifmmode {\mathrm{Me\kern -0.1em V}}\else
                   \textrm{Me\kern -0.1em V}\fi\,}%
\def\keV{\ifmmode {\mathrm{ke\kern -0.1em V}}\else
                   \textrm{ke\kern -0.1em V}\fi\,}%
\def\eV{\ifmmode  {\mathrm{e\kern -0.1em V}}\else
                   \textrm{e\kern -0.1em V}\fi\,}%
\let\tev=\TeV
\let\gev=\GeV
\let\mev=\MeV
\let\kev=\keV
\let\ev=\eV

\def\iab{\mbox{ab$^{-1}$}}
\def\ifb{\mbox{fb$^{-1}$}}
\def\ipb{\mbox{pb$^{-1}$}}
\def\inb{\mbox{nb$^{-1}$}}

\definecolor{Purple}{rgb}{0.6, 0.4, 0.8}
\newcommand{\VM}[1]{{ \color{orange} #1}}
\newcommand{\VMcom}[1]{{ \bf \color{Purple} #1}}
\newcommand{\VMrem}[1]{{ \color{gray} #1}}
\newcommand{\com}[1]{{ \color{red} #1}}
\newcommand{\myComment}[1]{}
\newcommand{\gd}[1]{\textcolor{green!70!black}{#1}}
\newcommand{\evr}[1]{\textcolor{blue}{#1}}

\newcommand{\HEPfit}{\texttt{HEPfit }}
\newcommand{\ttbar}{{\ensuremath{t\bar{t}}}}

\renewcommand{\thefootnote}{\arabic{footnote}}

\abstract{
In this contribution to the 2021 Snowmass community planning exercise that informs the American strategy for particle physics, we present the prospects for measurements of the top-quark couplings at future colliders. Projections are presented for the high luminosity phase of the Large Hadron Collider and a future Higgs/electroweak/top factory electron-positron collider. Results are presented for the expected bounds on Wilson coefficients of the relevant SMEFT operators from a global fit to the top physics sector.
}

\maketitle

\section{Introduction}

High-energy colliders provide a wealth of measurements of the rates, asymmetries and differential cross sections of a large number of processes. The Standard Model Effective Field Theory offers a systematic framework to order all these data and search for subtle patterns of deviations from the Standard Model. Fits of the dimension-six operator coefficients to projections for future experiments moreover offer an excellent benchmark to understand the constraining power of future data sets, without assuming ad-hoc scenarios for beyond-the-Standard-Model phenomena.

Since the discovery of the top quark, the Tevatron and LHC experiments have developed an extensive program of measurements of top-quark properties and interactions. Several groups have performed global fits to the SMEFT operator coefficients that affect the top-quark sector~\cite{Brivio:2019ius,Hartland:2019bjb,Buckley:2015lku} or the top/Higgs/electroweak sector~\cite{Ethier:2021bye,Ellis:2020unq}. 

Also bottom-quark production is experimentally distinguishable from other QCD processes that lead to final states with jets, thanks to precise vertex detectors and sophisticated flavour tagging algorithms. As the left-handed top and bottom quarks are part of the same weak-isospin doublet, and share a dependence on certain Wilson coefficients~\cite{Durieux:2019rbz, Durieux:2018tev}, we consider both third-generation quarks together in this work, including the bottom-quark operators that affect top-quark or otherwise relevant observables in our fit. 

The new collider facilities contemplated in the global road maps for particle physics (see Ref.~\cite{EuropeanStrategyGroup:2020pow} for the 2020 update of the European strategy) are expected to improve the measurement of top- and bottom-quark interaction rates and properties. The remaining LHC program, including the high-luminosity upgrade (HL-LHC), can provide precise rate measurements for rare top production processes and can extend the classical hadron-collider top production processes well into the boosted regime~\cite{Azzi:2019yne}. The main strength of an electron-positron collider operated above the top-quark pair production threshold is that it enables an ultra-precise characterisation of the top-quark electroweak (EW) interactions~\cite{Amjad:2015mma}. Finally, a new hadron or lepton collider that pushes the energy frontier to the 10~\TeV{} scale may open up a new kinematic regime, strongly enhancing the sensitivity to four-quark operators~\cite{Durieux:2018tev}. 

In this contribution, we study the impact of the HL-LHC and several future electron-positron collider scenarios on the top- and bottom-quark sectors of the SMEFT. All the fits presented here have been performed using the \HEPfit package \cite{DeBlas:2019ehy}.
We do not provide quantitative results for the potential of the highest-energy options (FCChh, muon collider, advanced electron-positron collider), but discuss some qualitative features. 

\section{SMEFT basis}

Effects of new physics in the couplings of the top quark can be described as effective
interactions of SM particles at energies below a new physics matching scale $\Lambda$. These effective interactions can be parameterised in terms of a set of Wilson coefficients $C_i$ of dimension-six operators $O_i$ in
the effective Lagrangian,

\begin{equation}
\mathcal{L}_\text{eff} = \mathcal{L}_\text{SM} +   \left( \frac{1}{\Lambda^2} \sum_i \mathcal{C}_i O_i + \text{h.c.} \right)  + \mathcal{O}\left(\Lambda^{-4} \right) ,
\end{equation}

\noindent
where, for this work, the sum runs over the operators shown in Tables~\ref{tab:WilsonDefined} and~\ref{tab:WilsonOperators} that involve top and bottom quarks, as described below, and which can be interpreted in terms of new physics mediators. This EFT preserves the  Lorentz and gauge symmetries of the SM, and operators with odd dimension are omitted since they violate baryon or lepton number.

These dimension-six operators contribute to the collider observables that we consider in this work. The leading-order contribution is given by the interference of the dimension-six operators with the SM, which generates linear terms in the operator coefficients divided by the square of the new physics scale $\Lambda$. Quadratic contributions are generated by squaring the amplitudes featuring a dimension-six operator insertion and lead to terms of order $\Lambda^{-4}$, the same order as the interference of the SM with the dimension-eight operators that we ignore in this work. Even if the known quadratic terms are often included in SMEFT fits~\cite{AguilarSaavedra:2018nen}, we opt for a more conservative approach here and include only the linear ones.
For the two-quark operators similar constraints could be obtained while using linear and linear plus quadratic terms, for most of the Wilson coefficients, as shown in Ref.~\cite{Miralles:2021dyw}.
For the four-quark operators the inclusion of quadratic terms helps to eliminate the blind directions since they reduce the possibility of having strong cancellations among the different contributions.
This effect can be observed when comparing the results of Ref.~\cite{Ellis:2020unq} (where only linear terms are included) with the results of Ref.~\cite{Ethier:2021bye} (where linear and quadratic terms are considered). 
Note also that, considering only linear terms, we lose sensitivity to the four-quark operators featuring a colour-singlet top current, since they do not interfere with the dominant QCD amplitudes for pair production.
We refer to Ref.~\cite{Brivio:2019ius} for a detailed study on the contributions of the top-quark operators to the observables included.

\begin{table}[tb]
\centering
\renewcommand{\arraystretch}{1.2}
\begin{tabular*}{\textwidth}{@{\extracolsep{\fill}}|c|c@{\quad}|c@{\quad}|c@{\quad}| } 
 \hline
 \multicolumn{4}{|c|}{\textbf{Coefficients fitted}} \\
 \hline
  \multirow{3}{*}{2-quark} & $C_{ t G}$  & $C_{\varphi Q}^3$  & $C_{\varphi Q}^- = C_{\varphi Q}^1-C_{\varphi Q}^3$  \\ 
    &   $C_{\varphi t}$ &  $C_{\varphi b}$  &    $C_{t Z} = c_W C_{t W}-s_W C_{t B}$   \\ 
 
    &  -- &  $C_{t \varphi }$ &   $C_{t W}$    \\ 
 \hline
 \hline
     \multirow{4}{*}{4-quark} &  $ C_{tu}^8 = \sum\limits_{\scaleto{i=1,2}{4pt}} 2C_{uu}^{(i33i)}$ &  $ C_{td}^8 = \sum\limits_{\scaleto{i=1,2,3}{4pt}} C_{ud}^{8(33ii)}$ &  $C_{Qq}^{1,8} = \sum\limits_{\scaleto{i=1,2}{4pt}} C_{qq}^{1(i33i)}+3C_{qq}^{3(i33i)}$  \\

        & $ C_{Qu}^{8} = \sum\limits_{\scaleto{i=1,2}{4pt}} C_{qu}^{8(33ii)}$ &  $ C_{Qd}^{8} =\sum\limits_{\scaleto{i=1,2,3}{4pt}} C_{qd}^{8(33ii)}$  &   $C_{Qq}^{3,8} =\sum\limits_{\scaleto{i=1,2}{4pt}} C_{qq}^{1(i33i)}-C_{qq}^{3(i33i)}$  \\

    &    -- &   --  &  $C_{tq}^{8} = \sum\limits_{\scaleto{i=1,2}{4pt}} C_{uq}^{8(ii33)}$  \\
 \hline
 \hline
 \multirow{3}{*}{\makecell{2-quark\\2-lepton}}  & $C_{eb}$ & $C_{et}$ & $ C_{ l Q}^+ = C_{lQ}^1+C_{lQ}^3$  \\
 
   & $C_{lb}$ &  $C_{lt}$ &  $C_{ l Q}^- = C_{lQ}^1-C_{lQ}^3$  \\
  
    &  -- & --  &   $C_{eQ}$ \\
   \hline
\end{tabular*}
\caption{Here we present the Wilson coefficients that have been fitted in our analysis in terms of those of Table~\ref{tab:WilsonOperators}. Those in  first block are related with the two-quark operators, those in the second block with the four-quark operators and the last block is related with the two-quark two-lepton operators.}
\label{tab:WilsonDefined}
\end{table}

\begin{table}[tb]
\centering
\renewcommand{\arraystretch}{1.2}
\begin{tabular*}{\textwidth}{@{\extracolsep{\fill}} |c|c@{\quad}||c@{\quad}|c@{\quad}| } 
 \hline
 \multicolumn{4}{|c|}{\textbf{Relevant operators}} \\
 \hline
 Coefficient & Operator & Coefficient & Operator \\ 
 \hline
$C_{\varphi Q}^1 $
		&   
		$\left(\bar{ Q} \gamma^\mu  Q \right) \left(\FDF[i] \right)  $ &$ C_{\varphi Q}^3$
		&$ 
		 \left(\bar{ Q} \tau^I \gamma^\mu  Q \right) \left(\FDFI[i] \right)$ \\ 
 $C_{\varphi t}$
		&
		$\left(\bar{ t}\gamma^\mu  t\right)\left(\FDF[i]\right) $& $C_{\varphi b}$
		& $
		\left( \bar{ b}\gamma^\mu  b \right)\left( \FDF[i]\right)$ \\ 
	     $C_{t\varphi}$
		&  
		$\left( \bar{ Q}  t \right)
		\left( \epsilon\varphi^* \; \varphi^\dagger\varphi \right) $ 
        &
        $C_{tG}$ & $\left(\bar{t} \sigma^{\mu \nu} T^{A} t \right)\left(\epsilon \varphi^* G_{\mu \nu}^{A} \right)$\\
	
	$C_{tW}$
		& 
		$\left(\bar{ Q}\tau^I\sigma^{\mu\nu}  t  \right)\left(\epsilon\varphi^* W_{\mu\nu}^I  \right) $&
		$ C_{tB} $ 
		&
		$	\left(\bar{ Q}\sigma^{\mu\nu}  t \right)
	 \left( \epsilon\varphi^* B_{\mu\nu} \right)$\\
 \hline
 \hline
 $C_{qq}^{1(ijkl)}$
		&  
		$(\bar{ q}_i\gamma^\mu  q_j) (\bar{  q}_k\gamma_\mu  q_l)$ &
 $C_{qq}^{3(ijkl)}$
		&  
		$(\bar{ q}_i\tau^I\gamma^\mu  q_j) (\bar{  q}_k\tau^I\gamma_\mu  q_l)$ \\
		
 $C_{uu}^{(ijkl)}$
		&  
		$(\bar{ u}_i\gamma^\mu  u_j) (\bar{  u}_k\gamma_\mu  u_l)$ &
 $C_{ud}^{8(ijkl)}$
		&  
		$(\bar{ u}_i\gamma^\mu T^A u_j) (\bar{  d}_k\gamma_\mu T^A d_l)$ \\

 $C_{qu}^{8(ijkl)}$
		&  
		$(\bar{ q}_i\gamma^\mu T^A q_j) (\bar{  u}_k\gamma_\mu T^A u_l)$  &
 $C_{qd}^{8(ijkl)}$
		&  
		$(\bar{ q}_i\gamma^\mu T^A q_j) (\bar{  d}_k\gamma_\mu T^A d_l)$ \\
 \hline
 \hline

	$C_{lQ}^1$
		&$  \left(\bar{ Q}\gamma_\mu  Q\right)	\left( \bar{ l}\gamma^\mu  l \right)$ &
	$C_{lQ}^3$
		& $ \left( \bar{ Q}\tau^I\gamma_\mu  Q \right)	\left(\bar{ l}\tau^I\gamma^\mu  l \right) $
	\\
	$C_{lt} $&$ \left( \bar{ t}\gamma_\mu  t\right) 	 \left(\bar{ l}\gamma^\mu  l \right)$ &
	$C_{lb} $&$  \left( \bar{ b}\gamma_\mu  b\right)		\left( \bar{ l}\gamma^\mu  l\right) $
	\\
	$C_{eQ} $&$ \left( \bar{ Q}\gamma_\mu  Q \right) 	\left(\bar{ e}\gamma^\mu  e \right)$&
	$C_{et} $& $ \left(\bar{ t}\gamma_\mu  t \right) 		\left( \bar{ e}\gamma^\mu  e \right)$
	\\
	$C_{eb} $& $ \left(\bar{ b}\gamma_\mu  b \right) 		\left( \bar{ e}\gamma^\mu  e \right)$& -- & -- \\
	\hline
\end{tabular*}
\caption{Here we show the most relevant operators whose linear combinations have been fitted in this work. The first block are two-quark operators, the second block are four-quark operators and the last block are two-quark two-lepton operators. In these operators $Q$ is the left-handed doublet of the two heaviest quarks, the Latin letters are flavour indices, $\tau^I$ are the Pauli matrices, $T^A=\lambda^A/2$ with $\lambda^A$ the Gell-Mann matrices.
}
\label{tab:WilsonOperators}
\end{table}

The number of operators involved in the SMEFT description is prohibitive if one adopts the most general flavour structure. We focus on the operator coefficients of the Warsaw basis~\cite{Grzadkowski:2010es} (see also Refs.~\cite{AguilarSaavedra:2008zc,Zhang:2010dr}) that involve at least one top quark, as well as the bottom-quark operators that affect the observables included in our study. Motivated by the minimal flavour violation ansatz, a $U(2)_q \otimes U(2)_u \otimes U(2)_d$ symmetry is imposed among the first two generations, as in the conventions proposed by the LHC Top Working Group~\cite{AguilarSaavedra:2018nen}.
The three lepton generations are treated independently but, since we only include two-quark two-lepton operators in $e^+e^-$ collider production processes, our analysis is only sensitive to the degrees of freedom corresponding to the electron.
We do not include the CP-violating imaginary parts of the Wilson coefficients, nor operators that lead to flavour-changing-neutral-current interactions.

The operator coefficients included in our analysis are listed in Table~\ref{tab:WilsonDefined} and the operators are defined in Table~\ref{tab:WilsonOperators}.
The selected sub-set of operators consists of three main blocks: the two-quark operators that modify top- and bottom-quark electroweak couplings and the \ttbar-gluon vertex, the four-quark operators of the type $q\bar{q}t\bar{t}$ (i.e.\ two light quarks and two heavy quarks) and the two-lepton-two-heavy-quark operators of the type $e^+e^-\ttbar$ and $e^+e^-b\bar{b}$. The four-quark operators are best probed at hadron colliders, while $e^+e^-$ colliders can provide better bounds on the two-lepton-two-quark operators. Both types of machines can provide bounds on the two-fermion operators and a direct comparison is possible for this set. As in Ref.~\cite{AguilarSaavedra:2018nen} we use the linear combinations $O_{\varphi Q}^- \equiv O_{\varphi Q}^1 - O_{\varphi Q}^3$ and $O_{qZ} \equiv - \sin \theta_W O_{qB} + \cos \theta_W O_{qW}$, and, likewise, $C_{lQ}^- \equiv C_{lQ}^1 - C_{lQ}^3$, as indicated in Table~\ref{tab:WilsonOperators}.

\section{Prospects for the HL-LHC}

In this section, we discuss the projections for the HL-LHC top physics measurements included in our fit.

The LHC is a top-quark factory that produces enormous samples of $t\bar{t}$ events and sizeable samples for rare top processes. The prospects for measurements at the LHC for the full 3000~\ifb{} expected after the high-luminosity phase of the LHC are based on an extrapolation from current (run 2) measurements. The measurements that form the basis for the HL-LHC projection are listed in Table~\ref{tab:measurements}, the inputs used for each observable can be found in App.~\ref{app:Inputs}.

The projections of the measurements of rare top-quark production processes are modelled on the S2 scenario used to predict the precision of Higgs coupling measurements in Ref.~\cite{Cepeda:2019klc}. This scenario envisages that many statistical and experimental uncertainties scale as $1/\sqrt{L_\text{int}}$, where $L_\text{int}$ is the integrated luminosity.
For the complete HL-LHC programme, experimental uncertainties are reduced by a factor 5. Theory and modelling uncertainties are divided by two, with respect to today's state of the art. This assumes N$^2$LO calculations will be achieved for rare associated production processes, and that Monte Carlo modelling can significantly be improved in the next decade.

To gain the maximal sensitivity to the EFT coefficients, differential measurements are included in our analysis. As in Ref.~\cite{Miralles:2021dyw}, for the $pp \rightarrow t\bar{t}Z$ and $pp \rightarrow t\bar{t}\gamma$ processes, differential measurements as a function of the $Z$-boson and photon $p_T$ are included, enhancing the sensitivity to $C_{tZ}$, in particular~\cite{Bylund:2016phk}.

\begin{table*}[tb]
\centering
\resizebox{\textwidth}{!}{%
\begin{tabular}{|l|c|c|c|c|c|c|}
\hline
Process & Observable & $\sqrt{s}$  & $L_\text{int}$  & Experiment & SM  & Ref.\\ \hline
$pp \rightarrow \ttbar $ & $d\sigma/dm_\ttbar$ (15+3 bins) & 13 TeV & 140~\ifb & CMS & \cite{Czakon:2013goa} & \cite{CMS:2021vhb} \\
$pp \rightarrow \ttbar $ & $dA_C/dm_\ttbar$ (4+2 bins) & 13 TeV & 140~\ifb & ATLAS & \cite{Czakon:2013goa} & \cite{ATLAS-CONF-2019-026} \\
$p p \rightarrow t \bar{t} H+ tHq$ & $\sigma$ & 13 TeV & 140~\ifb & ATLAS & 
\cite{deFlorian:2016spz} &  \cite{ATLAS:2020qdt} \\ 
$p p \rightarrow t \bar{t} Z$ &  $d\sigma/dp_T^Z$ (7 bins) & 13 TeV &  140~\ifb  & ATLAS &
\cite{Broggio:2019ewu} & \cite{ATLAS:2020cxf} \\ 
$p p \rightarrow t \bar{t} \gamma$ & $d\sigma/dp_T^\gamma$ (11 bins) & 13 TeV & 140~\ifb & ATLAS &
\cite{Bevilacqua:2018woc,Bevilacqua:2018dny} & \cite{Aad:2020axn}  \\ 
$p p \rightarrow tZq$ & $\sigma$ & 13 TeV & 77.4~\ifb{}  & CMS &
 \cite{Sirunyan:2017nbr} & \cite{Sirunyan:2018zgs} \\ 
$p p \rightarrow t\gamma q$ & $\sigma$ & 13 TeV & 36~\ifb{}  & CMS &
 \cite{Sirunyan:2018bsr} & \cite{Sirunyan:2018bsr} \\
 $p p \rightarrow t \bar{t} W$ & $\sigma$ & 13 TeV & 36~\ifb  & CMS &
\cite{deFlorian:2016spz,Frederix:2017wme} &  \cite{Sirunyan:2017uzs} \\
 $p p \rightarrow t\bar{b}$ (s-ch) & $\sigma$ & 8~\tev{} & 20~\ifb{} & LHC &
\cite{Aliev:2010zk,Kant:2014oha} & \cite{Aaboud:2019pkc} \\ 
$p p \rightarrow tW$ & $\sigma$ & 8~\TeV & 20~\ifb{}  & LHC &
 \cite{Kidonakis:2010ux} &  \cite{Aaboud:2019pkc}   \\ 
$p p \rightarrow tq$ (t-ch) & $\sigma$ & 8~\tev{} & 20~\ifb{} & LHC &
\cite{Aliev:2010zk,Kant:2014oha} & \cite{Aaboud:2019pkc} \\ 
$t \rightarrow Wb $ & $F_0$, $F_L$  & 8~\tev{} & 20~\ifb{} & LHC &
\cite{PhysRevD.81.111503}  & \cite{Aad:2020jvx} \\
$p\bar{p} \rightarrow t\bar{b}$ (s-ch) & $\sigma$ & 1.96~\tev & 9.7~\ifb & Tevatron & \cite{PhysRevD.81.054028} & \cite{CDF:2014uma} \\
$e^{-} e^{+} \rightarrow b \bar{b} $ & $R_{b}$ ,  $A_{FBLR}^{bb}$ & $\sim$ 91~\GeV & 202.1~\ipb  & 
LEP/SLD &
 - & \cite{ALEPH:2005ab}  \\ \hline
\end{tabular}%
}
\caption{Measurements included in the EFT fit of the top-quark electroweak sector. For each measurement, the process, the observable, the centre-of-mass energy, the integrated luminosity and the experiment/collider are given. The last two columns list the references for the predictions and measurements that are included in the fit. LHC refers to the combination of ATLAS and CMS measurements. In a similar way, Tevatron refers to the combination of CDF and D0 results, and LEP/SLD to different experiments from those two accelerators.}
\label{tab:measurements}
\end{table*}

For the top-quark pair production process, statistics is abundant and measurements in the bulk already reach a precision of a few \%. The inclusive cross section measurement is currently limited to a 2\% uncertainty~\cite{ATLAS:2016zet} by the uncertainty on the integrated luminosity. This uncertainty is expected to be reduced to approximately 1\% at the HL-LHC~\cite{Azzi:2019yne}. Currently, theory uncertainties of the N$^2$LO calculation are at the level of 3--4\% for the inclusive cross section~\cite{Czakon:2013goa}. These might be reduced to roughly half with the calculation of the N$^3$LO corrections and the improvement of the proton PDFs. Even in that case, theory uncertainties are likely to remain the limiting factor.

The $t\bar{t}$ charge asymmetry is a subtle effect at the LHC, but it  brings important information to EFT fits~\cite{Zhang:2012cd}. As a ratio, it can be precisely predicted~\cite{Czakon:2017lgo}. Modelling uncertainties play an important role~\cite{ATLAS-CONF-2019-026} and are likely to limit future progress in the inclusive measurement. Therefore, a less aggressive scenario is adopted, where all experimental systematic uncertainties are improved by a factor 1/2 and only the statistical uncertainty scales with $1/\sqrt{L_\text{int}}$.

\begin{figure}[h!]
\includegraphics[trim=115 55 175 130, clip, width=1.0\columnwidth]{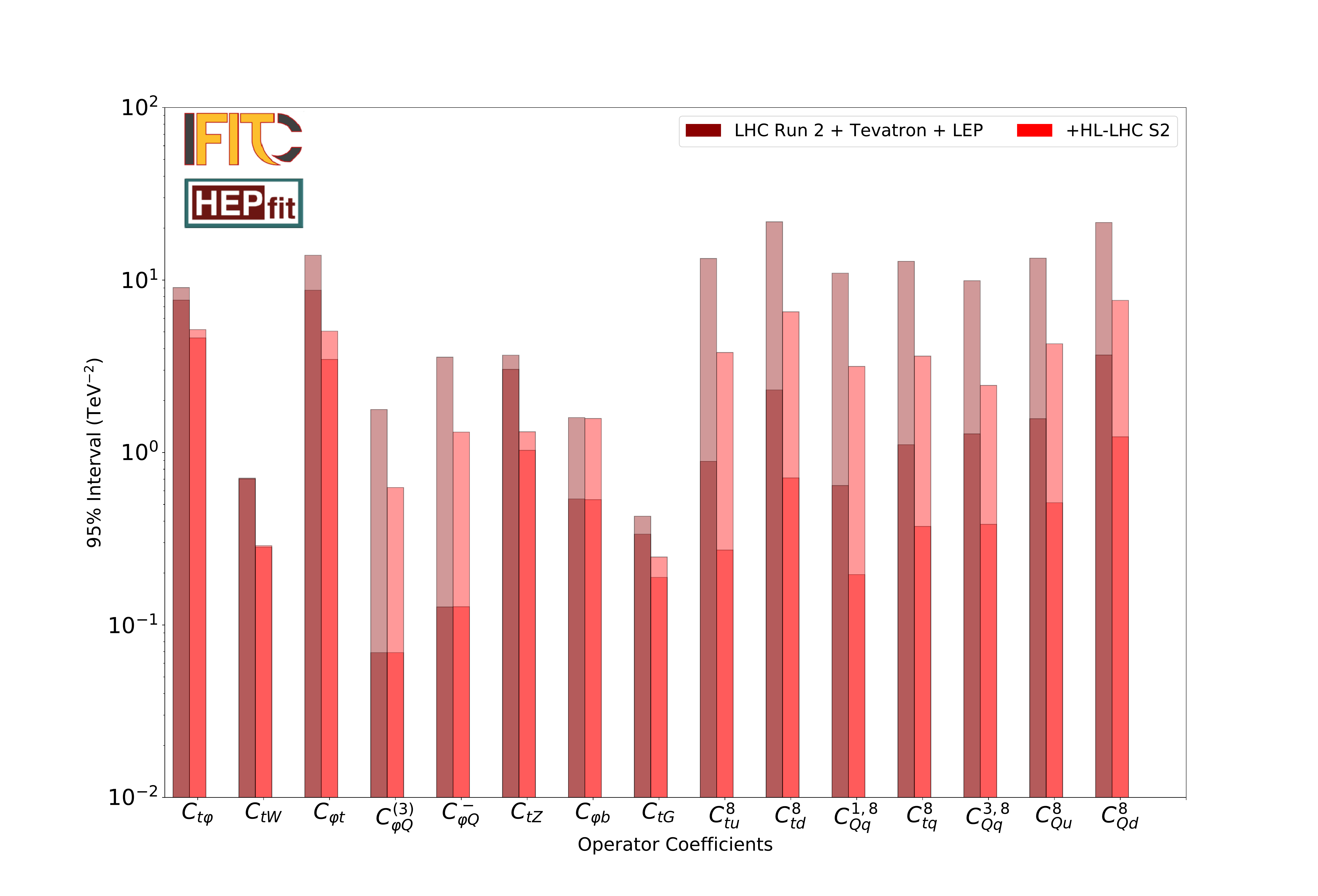}
\caption{\label{fig:hllhc_projection}%
The 95\% probability bounds on the Wilson coefficients for dimension-six operators that affect the top-quark production and decay measurements listed in Table~\ref{tab:measurements} after run 2 of the LHC (in dark red) and prospects for the bounds expected after completion of the complete LHC program, including the high-luminosity stage (in light red). Only linear terms proportional to $\Lambda^{-2}$ are taken into account in the dependence of the observables on the Wilson coefficients. The individual bounds obtained from a single-parameter fit are shown as solid bars, while the global or marginalised bounds obtained fitting all Wilson coefficients at once are indicated by the full bars (shaded region in each bar). The correlations between the Wilson coefficients obtained in the global fit can be found in App.~\ref{app:correlation_matrices}.
}
\end{figure}

For top-quark pair production, differential measurements of the cross section~\cite{CMS:2021vhb} and the charge asymmetry~\cite{ATLAS-CONF-2019-026} as a function of the invariant mass of the $t\bar{t}$ system are considered. A promising avenue for progress is the boosted regime, where the sensitivity to four-fermion operators increases considerably~\cite{Rosello:2015sck}. Measurements of the cross section and charge asymmetry for $t\bar{t}$ systems produced at large invariant mass already play an important role in the constraints on the four-fermion operators and their weight will increase if measurements on bulk $t\bar{t}$ are limited by experimental or theoretical systematic uncertainties. To take maximal advantage of this potential, the range of the projections is extended further into the high-$m_\ttbar$ tail than the current run 2 measurements. More details in the binning that has been considered can be found in App.~\ref{app:differential}.

The 95\% probability bounds from a fit to the current data are shown as the dark red bars in Fig.~\ref{fig:hllhc_projection}. The data include LHC run 2 and run 1 measurements and several legacy results from the Tevatron, LEP, and SLD experiments. The light red bars present the extrapolation to the complete HL-LHC program, with an integrated luminosity of 3~\iab{}.

Across the board, the HL-LHC program is expected to improve the bounds by a factor of two to four with respect to the current run 2 limits, both for individual bounds and global fit results. Exceptions are the individual bounds on $C_{\varphi Q}^-$ and $C_{\varphi Q}^3$, that continue to depend on the $Zb\bar{b}$ measurements at the $Z$-pole. 

Generally, the progress envisaged in the S2 scenario is limited by the theory and modelling uncertainties, while statistical and experimental uncertainties  are expected to be sub-dominant in nearly all measurements in the S2 scenario. Therefore, improving the accuracy of fixed-order predictions beyond the factor two envisaged in the S2 scenario, which already assumes significant advances in the theoretical calculations, will lead to a direct improvement of the sensitivity. This will, however, likely require N$^3$LO precision for $2 \rightarrow 3$ processes with top quarks in the final state.

The boosted regime is indeed confirmed as one of the keys to improving bounds on the operators that affect the top-quark pair production process. In particular, the high-$m_\ttbar$ tail of the top-quark pair production measurements provides a significant reduction in the allowed regions of the four-quark operators, which shrink by a factor between two and five (depending on the operator) thanks to the enhanced sensitivity in this regime and the more pronounced improvement in the measurement. This effect is present even in a fit that only includes the linear (${\cal O} (\Lambda^{-2}$)) terms in the parameterisation of the EFT dependence. 

The marginalised bounds on the four-fermion operators remain an order of magnitude worse than the individual bounds after the HL-LHC, even if both individual and global bounds improve considerably. This is due to unresolved correlations between the coefficients. The same feature is observed in recent fits to the top sector of the SMEFT~\cite{Brivio:2019ius,Hartland:2019bjb} and in global Higgs/EW/top fits~\cite{Ethier:2021bye,Ellis:2020unq}. Stricter limits can be obtained if the dimension-six-squared terms proportional to $\Lambda^{-4}$ are included in the fit~\cite{Ethier:2021bye}.

Two-quark two-lepton operators, omitted in this section, can also be probed at the LHC.
Dedicated signal regions, for instance with off-$Z$-peak dilepton invariant masses in $pp\to t\bar{t}\ell^+\ell^-$~\cite{Durieux:2014xla, Chala:2018agk, CMS:2020lrr}, would increase their sensitivity.

\section{Prospects for \texorpdfstring{$e^+e^-$}{e+e-} colliders}

In this section, we study the impact of precision measurements in $e^+e^- \rightarrow  b\bar{b} $ and $e^+e^- \rightarrow t\bar{t}$ production, using operating scenarios of the main circular and linear collider concepts.

Prospects for the $e^+e^- \rightarrow b\bar{b}$ process are included that are based on the full-simulation studies of the ILD concept~\cite{Okugawa:2019ycm} at $\sqrt{s} = $ 250~\GeV. The prospects are based on realistic estimates of efficiency and acceptance, including the signal losses required to ensure a robust calibration of the flavour tagging efficiency. The statistical uncertainties on the measurements of the cross section and forward-backward asymmetry are complemented by polarisation and flavour-tagging systematics. For the $Z$-pole runs we use the projections for $R_b$ and $A_{FB}$ provided by the FCCee and CEPC projects for their ``TeraZ'' runs at the $Z$-pole.

\begin{figure}[tb]
\includegraphics[trim=115 55 175 145, clip, width=\textwidth]{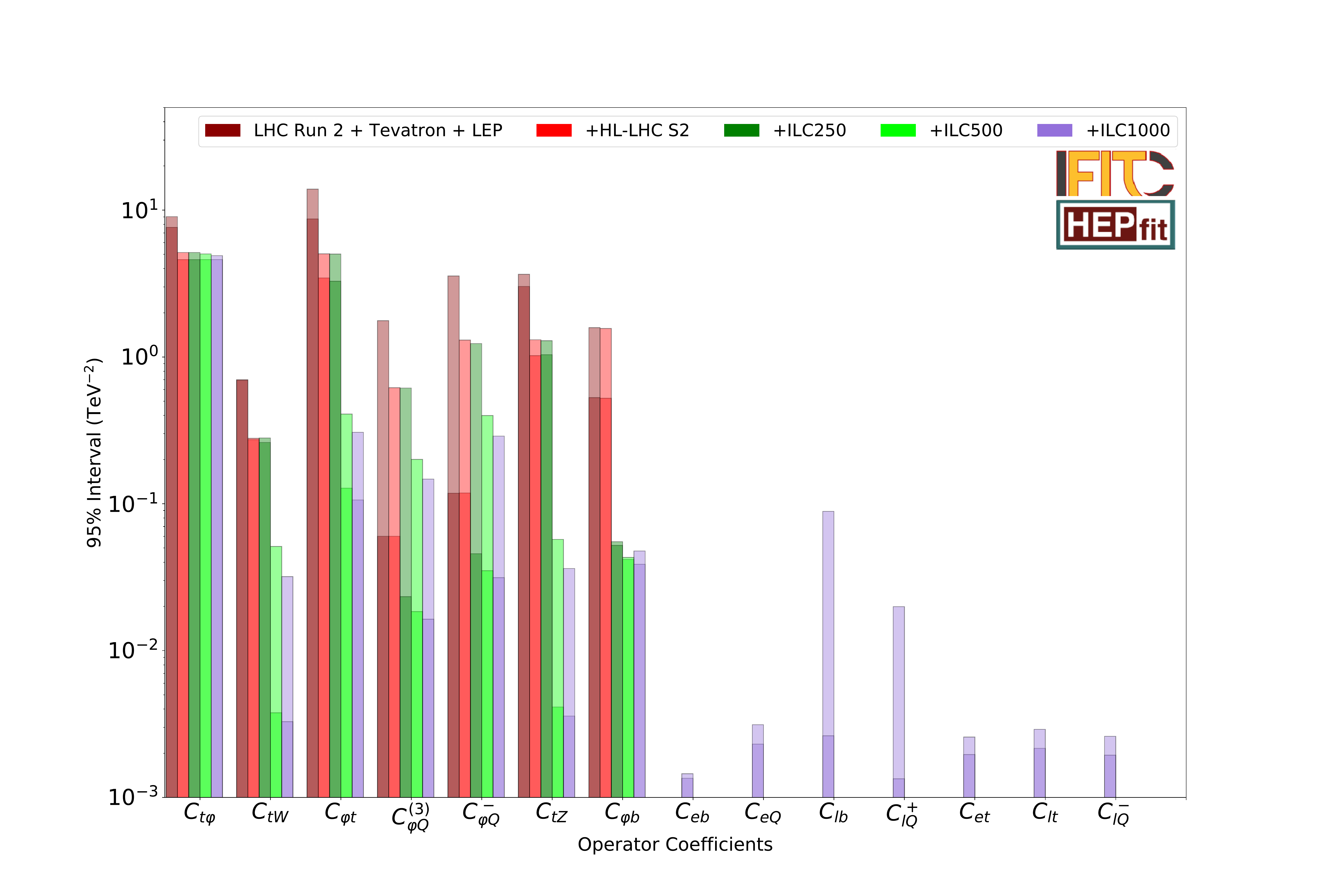}
\caption{\label{fig:hllhc_ILC_projection}%
Comparison of current LHC constraints with HL-LHC ones, and those deriving from ILC runs at 250, 500 and 1000~GeV.
The limits on the $q\bar{q}t\bar{t}$ and $C_{tG}$ coefficients are not shown, since the $e^+e^-$ collider measurements considered are not sensitive to them, but all operators are included in the global fit.
The improvement expected from the HL-LHC on these coefficients is shown in Fig.~\ref{fig:hllhc_projection}. The solid bars provide the individual limits of the single-parameter fit and the shaded ones the marginalised limits of the global fit. The correlations between the Wilson coefficients obtained in the global fit can be found in App.~\ref{app:correlation_matrices}.}
\end{figure}

The $e^+e^- \rightarrow t\bar{t}$ process opens up for centre-of-mass energies that exceed twice the top mass (i.e. $\sqrt{s} \gtrsim$ 350~\GeV) and probes the electroweak couplings of the top quark at tree-level.
Data taken with different beam polarisations at linear colliders can be used to distinguish the photon and $Z$-boson couplings~\cite{Amjad:2013tlv,Amjad:2015mma,Durieux:2018tev,CLICdp:2018esa}.
At circular colliders, a measurement of the final state polarisation using the semi-leptonically decaying top quarks can also be used to separate the two contributions~\cite{Janot:2015yza}.
We base our prospects on the study of statistically optimal observables defined at leading order on the $e^+e^- \to t\bar{t} \rightarrow WbWb$ differential distribution~\cite{Durieux:2018tev}.
This $WbWb$ final state also receives contribution from single top production which become sizeable at high centre-of-mass energies.
Realistic acceptance, identification and reconstruction efficiencies are estimated from full-simulation studies for the ILC and CLIC in Ref.~\cite{Amjad:2013tlv,Abramowicz:2016zbo}.
Since they were performed only for sub-set of centre-of-mass energies and beam polarisations, overall efficiency factors are extrapolated as a functions of the centre-of-mass energy.
They drop significantly for the TeV centre-of-mass energies of ILC and CLIC since a degradation of top-selection and flavour-tagging capabilities is expected in this regime.

In Fig.~\ref{fig:hllhc_ILC_projection}, the impact of runs at different centre-of-mass energies is illustrated.  The current bounds in brown are compared to HL-LHC ones in red.
The subsequent bars add data at $\sqrt{s}= 250~\gev$, 500~\gev{} and 1~\tev.
The beam polarisations and integrated luminosities of the different ILC stages are summarised in Table~\ref{tab:epem_setup}.
Only the electroweak operators are presented, as the $e^+e^-$ data have the strongest impact there, but results corresponds to a global analysis, including also the $q\bar{q}t\bar{t}$ operators and $C_{t G}$.

The dark green bar shows that the ``Higgs factory" run improves the bounds on bottom-quark operators, including $C_{\varphi Q}^3$ and $C_{\varphi Q}^-$ and $C_{\varphi b}$. The improvement is especially pronounced for the individual bounds. As expected, data above the top-quark pair production threshold is required to improve the bounds on the top-quark operators. 

Runs at two different centre-of-mass energies above the top-quark pair production threshold are required to disentangle the $e^+e^-t\bar{t}$ operator coefficients from the two-fermion operator coefficients~\cite{Durieux:2018tev}.
The two sets of operators have very different scaling with energy: the sensitivity to four-fermion operators grows quadratically, while it is constant or grows only linearly for two-fermion operators.
In a fit to data taken at a single centre of mass, linear combinations of their coefficients remain degenerate and form blind directions.
The combination of runs at two different centre-of-mass energies effectively disentangles them and provides global fit constraints close to the individual bounds

\begin{figure}[tb]
\includegraphics[trim=115 55 175 145, clip, width=1.0\columnwidth]{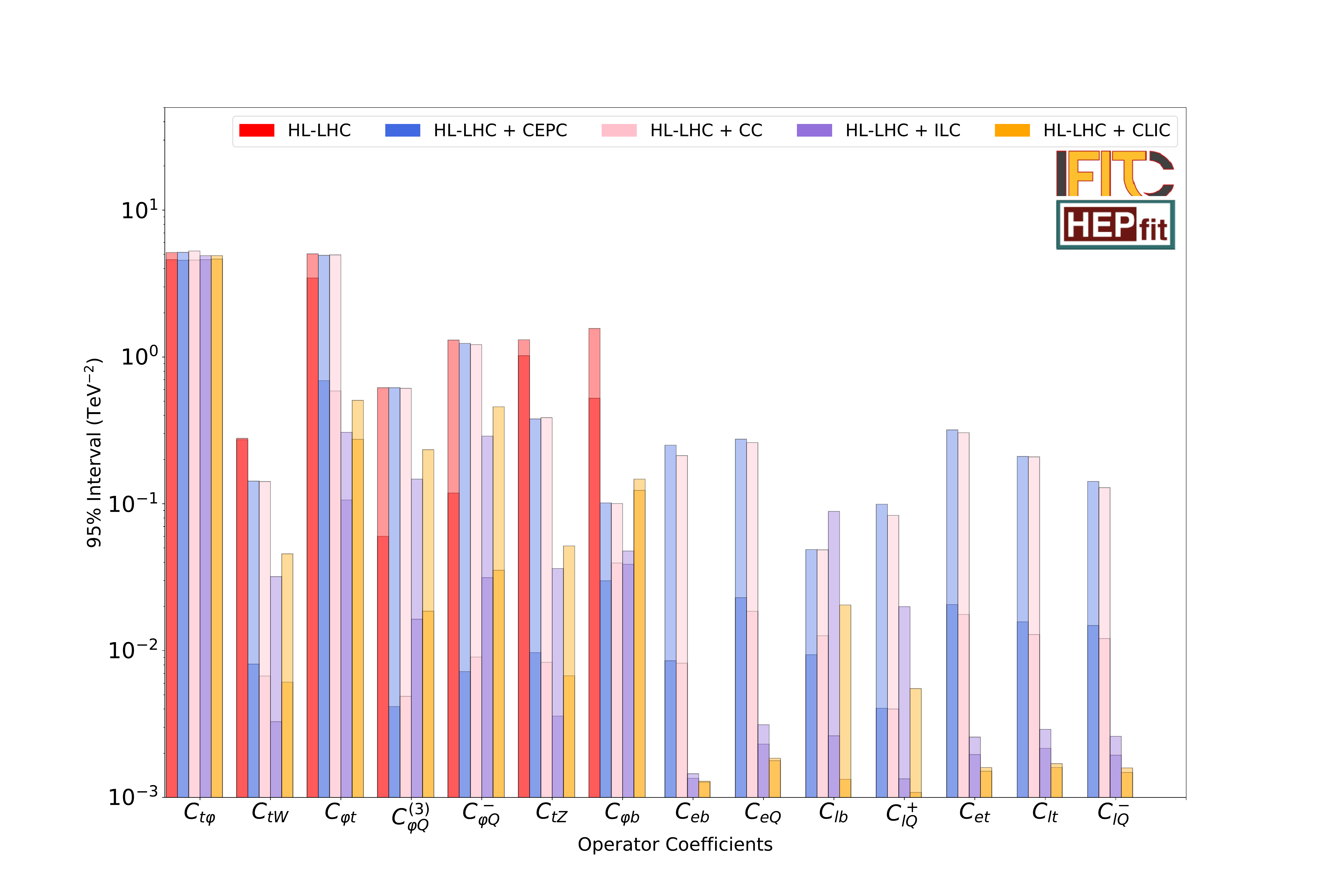}
\caption{\label{fig:CEPC_CC_ILC_CLIC_projection}%
Comparison of the constraints expected from a combination of HL-LHC and lepton collider data.
The limits on the $q\bar{q}t\bar{t}$ and $C_{tG}$ coefficients are not shown, since the $e^+e^-$ collider measurements considered are not sensitive to them, but all operators are included in the global fit.
The improvement expected from the HL-LHC on these coefficients is shown in Fig.~\ref{fig:hllhc_projection}. The solid bars provide the individual limits of the single-parameter fit and the shaded ones the marginalised limits of the global fit. The correlations between the Wilson coefficients obtained in the global fit can be found in App.~\ref{app:correlation_matrices}. }
\end{figure}

Several further processes are accessible to $e^+e^-$ colliders, but have not been taken into account in this study.
The top-quark Yukawa coupling can be determined through the tree-level dependence of the associated $e^+e^- \rightarrow t\bar{t}H$ production process. This requires runs with a centre-of-mass energy above 500--550~\GeV.
At linear colliders, where the luminosity grows with energy, there is a broad plateau up to about 1.5~\TeV{} where $e^+e^- \rightarrow t\bar{t}H$ is accessible.
Based on full-simulation studies of Ref.~\cite{Price:2014oca} we expect an improvement in the constraint on $C_{t\varphi}$ by a factor two with respect to the HL-LHC.

\begin{table}[tb]
    \centering
    \renewcommand{\arraystretch}{1.1}
    \begin{tabular*}{\textwidth}{@{\extracolsep{\fill}}|c|c@{\quad}|c@{\quad}|c@{\quad}|c@{\quad}|}
    \hline
       Machine & Polarisation & Energy & Luminosity & Reference\\\hline
        \multirow{3}{*}{ILC} & \multirow{3}{*}{P($e^+$, $e^-$):$(\pm30\%,\,\mp80\%)$}  & 250 GeV & 2 \iab{} & \multirow{3}{*}{\cite{AlexanderAryshev:2022pkx}} \\
         &  & 500 GeV & 4 \iab{} & \\
         &  & 1 TeV & 8 \iab{} & \\\hline
        \multirow{3}{*}{CLIC} & \multirow{3}{*}{P($e^+$, $e^-$):$(0\%,\,\pm80\%)$} &  380 GeV & 1 \iab{} & \multirow{3}{*}{\cite{Robson:2018zje}}\\
         &  & 1.4 TeV & 2.5 \iab{} & \\
         &  & 3 TeV & 5 \iab{} & \\\hline
         \multirow{4}{*}{FCC-$ee$} & \multirow{4}{*}{Unpolarised} &  Z-pole & 150 \iab{} & \multirow{4}{*}{\cite{Bernardi:2022hny}}\\
         &  & 240 GeV & 5 \iab{} & \\
         &  & 350 GeV & 0.2 \iab{} & \\
         &  & 365 GeV & 1.5 \iab{} & \\\hline
          \multirow{4}{*}{CEPC} & \multirow{4}{*}{Unpolarised} &  Z-pole & 57.5 \iab{} & \multirow{4}{*}{\cite{Bernardi:2022hny}}\\
         &  & 240 GeV & 20 \iab{} & \\
         &  & 350 GeV & 0.2 \iab{} & \\
         &  & 360 GeV & 1 \iab{} & \\\hline
    \end{tabular*}
    \caption{Here we show the different working configurations for the future $e^+e^-$ colliders.}
    \label{tab:epem_setup}
\end{table}

In Fig.~\ref{fig:CEPC_CC_ILC_CLIC_projection}, we compare the bounds expected from the HL-LHC and from the final stages of the CEPC, FCC-$ee$, ILC and CLIC.
The centre-of-mass energies, integrated luminosities and beam polarisations envisaged for each of these projects are given in Table~\ref{tab:epem_setup}. 
The circular colliders (FCC-$ee$ and CECP) operated at and slightly above the $t\bar{t}$ threshold are expected to improve constraints on the bottom- and top-operators by factors 5 and 2 for some two-fermion operators.
Indeed, their ``TeraZ'' runs provide very competitive bounds (individual ones, in particular) on two-fermion bottom-operator coefficients.
Their constraining power on four-fermion operators is, however, limited by the energy reach.
Since, at these colliders, the two runs above the $t\bar{t}$-threshold are very close the two-fermion and four-fermion operators are harder to disentangle.
The global limits remain significantly above the individual bounds.

The linear colliders (ILC and CLIC), operated at two centre-of-mass energies above the $t\bar{t}$ threshold, can provide very tight bounds on all operators.
The bounds on four-fermion operators take advantage of the energy-growing sensitivity and become very competitive if $e^+e^-$ collision data at a centre-of-mass energy greater than 1 TeV is available.
The ILC1000 and CLIC3000 bounds of $\mathcal{O}(10^{-3})$ on the $e^+e^-t\bar{t}$ operators are by far the tightest top-sector SMEFT constraints that can be achieved at any future collider considered in this work.\footnote{A muon collider or advanced linear collider have the potential to improve these bounds further, but quantitative projections for integrated luminosity and experimental performance are currently not available.}

\section{Pushing the energy frontier}

Several projects have been defined that extend the energy of colliders well beyond the TeV scale. Collisions at a centre-of-mass energy of 10~\TeV{} and beyond could be achieved
at a large (100 km circumference) hadron collider~\cite{FCC:2018vvp, CEPC-SPPCStudyGroup:2015csa},
at a linear electron-positron collider implementing novel accelerating techniques~\cite{Cros:2019tns},
or at compact circular muon colliders~\cite{InternationalMuonCollider:2022qki}.
The potential of these machines for the SMEFT fit lies mainly in the energy-growing sensitivity to new physics.
In the top- and bottom-quark sectors of the SMEFT, the sensitivity to four-fermion operators shows a strong increase~\cite{Aguilar-Saavedra:2014iga,Rosello:2015sck,Englert:2016aei,Durieux:2018tev}. For a given measurement precision, bounds derived in higher-energy collisions are therefore much stronger than those derived from measurements at lower energy. 

We illustrate the increased sensitivity with the dependence of the differential cross section at high $m_{\ttbar}$ to $C_{tu}^8$ and $C_{tG}$. At the LHC, the cross section measurement in the boosted regime (with $m_\ttbar > $ 1.4~\TeV), yields the following relation with the Wilson coefficient: 
\begin{equation}
\sigma(m_\ttbar > 1.4~\tev) = 1.8~\mathrm{pb} \times [1 + 0.3  \cdot C_{tG} + 0.1 \cdot C_{tG}^2 + 0.1 \cdot C_{tu}^8 + 0.3 \cdot (C_{tu}^8)^2 + ... ] 
\end{equation}
A 100~\TeV pp collider has a seven times larger energy reach and one could envisage a measurement with $m_{\ttbar} > $ 10~\tev{} that would have the following dependence:
\begin{equation}
\sigma(m_\ttbar > 10~\tev) = 0.1~\mathrm{pb} \times [1+ 0.3 \cdot C_{tG} + 1.8 \cdot C_{tG}^2 + 3 \cdot C_{tu}^8 + 256 \cdot (C_{tu}^8)^2 + ...] 
\end{equation}
The increase in the factors that multiply the Wilson coefficients is very clear for the quadratic term of $C_{tG}$ and for both the linear and quadratic terms in $C_{tu}^8$ (and similarly for the other four-fermion operator coefficients).\footnote{Note that here we are showing the impact on linear and quadratic terms since many studies include both contributions. We remark that in the quantitative analysis presented in this work only linear terms are considered, as explained above.}

We therefore expect that FCChh and SPPC measurements in the 10~\tev{} regime, with precision comparable to that of current boosted measurements at the LHC, could provide bounds that are a factor 20 sharper than the HL-LHC prospects, bringing the constraints down from ${\cal O}(1~\tev^{-2})$ to ${\cal O}(0.1~\tev^{-2})$.
This, of course, requires that techniques be developed to efficiently trigger, select and reconstruct events with highly boosted top quarks~\cite{Kogler:2018hem, Nachman:2022emq} and that the experimental response and Monte Carlo modelling be controlled to a similar level. 

\section{Conclusion}

New energy-frontier colliders are expected to provide an important push for top- and bottom-quark physics. We assess this potential in the framework of the Standard Model Effective Field Theory, performing fits to measurement projections in several scenarios. 

For the study of rare top production processes at the High-Luminosity phase of the LHC, we adopt the S2 scenario that is also used for Higgs measurements~\cite{Cepeda:2019klc}.
It envisages a strong $(\propto 1/\sqrt{L_\text{int}})$ reduction of experimental uncertainties and a reduction by a factor two for theory and modelling uncertainties.
For the very abundant top-quark pair production, a more conservative scenario is adopted.
Across the board, the HL-LHC is expected to improve bounds on Wilson coefficients by a factor three with respect to the current results that are dominated by LHC run 2. Global bounds on two-fermion operator coefficients are expected to range between 0.03 \tev$^{-2}$ and 6 \tev$^{-2}$, while four-fermion operator coefficients remain of order 1 due to strong degeneracies. 

A new electron-positron collider can provide very tight bounds on bottom-quark operators and --- if operated above the $\ttbar$ production threshold --- on the top-quark ones.
Constraints on the two-fermion operator coefficients that affect the top- and bottom-quark electroweak couplings improve by up to two orders of magnitude, reaching order 0.1~\tev$^{-2}$ for most coefficients.
With runs at several energies above the \ttbar{} production threshold the entire sector can be tightly constrained.
Operation at TeV centre-of-mass energies would yield bounds of order $10^{-3}$ \tev$^{-2}$ on the $e^+e^-t\bar{t}$ operator coefficients, making them the most precisely constrained operators in the top sector of the SMEFT.
The $Z$-pole run offers the best bounds on the two-fermion bottom-quark operators.

An energy-frontier proton collider is expected to further improve the bounds on four-fermion operators involving a pair of top quarks.
While no systematic and quantitative projections have been released, the increased sensitivity to $q\bar{q}t\bar{t}$ operators in the highly boosted regime offers the potential for an order-of-magnitude improvement with respect to the HL-LHC projections at a 100~TeV proton collider.

\FloatBarrier

\section*{Acknowledgments}

The work of VM is supported by the Italian Ministry of Research (MUR) under the grant PRIN20172LNEEZ. EV has received funding from the European Research Council (ERC) under the European Union’s Horizon 2020 research and innovation programme(Grant agreement No. 949451) and from a Royal Society University Research Fellowship through grant URF/R1/201553. RP acknowledges the support from the Leverhulme Trust and the Isaac Newton Trust,
as well as the use of the DiRAC Cumulus HPC facility under Grant No.\ PPSP226. The work of MV is supported by the projects No. PGC2018-094856-B-100 (MICINN), PROMETEO-2018/060
(Generalitat Valenciana) and ILINKB20065 (CSIC). The work of MMLL is supported by the projects SEJI-2020/037 (Generalitat Valenciana) and RYC2019-028510-I (MICINN). The work of AGC is supported by the program JAEICU-21-IFIC-6 (CSIC).

\appendix

\section{Appendix: Inputs included in the fits}

\label{app:Inputs}

\subsection*{LHC and HL-LHC inputs}

\begin{table}[h!]
\centering
\hspace*{-0.8 cm}\resizebox{1.15\textwidth}{!}{%
\begin{tabular}{|l|c|c|c|c|c|c|c|c|c|c|}
\hline
\multirow{3}{*}{$p p \rightarrow t \bar{t} $} &\multirow{3}{*}{ Measured ($\text{fb}\cdot\text{GeV}^{-1}$)}& \multirow{3}{*}{SM ($\text{fb}\cdot\text{GeV}^{-1}$)} & \multicolumn{4}{c|}{LHC Unc. ($\text{fb}\cdot\text{GeV}^{-1}$)}   &  \multicolumn{4}{c|}{HL-LHC Unc. ($\text{fb}\cdot\text{GeV}^{-1}$)}   \\ \cline{4-11}
& && \multirow{2}{*}{theo.} & \multicolumn{3}{c|}{exp.}  & \multirow{2}{*}{theo.} & \multicolumn{3}{c|}{exp.} \\ \cline{5-7} \cline{9-11}
& && & stat. & sys. + mod. & tot. & & stat. & sys. + mod. & tot.\\ \hline
$m_{t\bar{t}}:$ (250-400)     & 344.2  & 328.09  & 42.68     & 1.604    & 13.19  & 13.29  & 21.34    & 0.3465   & 6.600    & 6.60  \\ 
$m_{t\bar{t}}:$ (400-480)     & 871.7  & 883.38  & 111.28    & 4.784    & 41.29  & 41.57  & 55.64    & 1.0335   & 20.65    & 20.67  \\ 
$m_{t\bar{t}}:$ (480-560)     & 542.8  & 545.13  & 72.26     & 4.737    & 23.07  & 23.55  & 36.13    & 1.0233   & 11.54    & 11.58 \\
$m_{t\bar{t}}:$ (560-640)     & 315.3  & 318.36  & 46.58     & 4.561    & 14.22  & 14.93  & 23.29    & 0.9853   & 7.110    & 7.18 \\ 
$m_{t\bar{t}}:$ (640-720)     & 180.9  & 182.33  & 24.41     & 4.331    & 8.395  & 9.446  & 12.21    & 0.9356   & 4.198    & 4.30 \\
$m_{t\bar{t}}:$ (720-800)     & 109.6  & 109.27  & 18.79     & 3.748    & 5.468  & 6.629  & 9.395    & 0.8097   & 2.734    & 2.85 \\
$m_{t\bar{t}}:$ (800-900)     & 62.51  & 64.83   & 10.94     & 2.253    & 3.026  & 3.772  & 5.470    & 0.4867   & 1.513    & 1.59 \\
$m_{t\bar{t}}:$ (900-1000)    & 36.97  & 36.59   & 6.877     & 1.550    & 1.988  & 2.521  & 3.439    & 0.3348   & 0.994    & 1.05 \\
$m_{t\bar{t}}:$ (1000-1150)   & 17.84  & 19.15   & 3.904     & 0.6554   & 0.9134 & 1.124  & 1.952    & 0.1416   & 0.4567   & 0.48 \\
$m_{t\bar{t}}:$ (1150-1300)   & 9.005  & 9.060   & 2.020     & 0.4129   & 0.5086 & 0.6551 & 1.010    & 0.08920  & 0.2543   & 0.27 \\
$m_{t\bar{t}}:$ (1300-1500)   & 3.815  & 4.256   & 1.113     & 0.2151   & 0.2024 & 0.2954 & 0.5565   & 0.04647  & 0.1012   & 0.11 \\
$m_{t\bar{t}}:$ (1500-1700)   & 1.769  & 1.777   & 0.5259    & 0.1441   & 0.1108 & 0.1817 & 0.2630   & 0.03113  & 0.0554   & 0.064 \\
$m_{t\bar{t}}:$ (1700-2000)   & 0.6941 & 0.6806  & 0.2203    & 0.06014  & 0.0397 & 0.0720 & 0.1101   & 0.01298  & 0.01985  & 0.0237 \\
$m_{t\bar{t}}:$ (2000-2300)   & 0.1905 & 0.23647 & 0.0994    & 0.03459  & 0.0170 & 0.0386 & 0.0497   & 0.007474 & 0.00850  & 0.0113 \\
$m_{t\bar{t}}:$ (2300-2600)$^*$ &   --   & 0.12738 & 0.0640    &    --    &   --   &   --   & 0.0320   & 0.005416 & 0.00300  & 0.0072 \\ 
$m_{t\bar{t}}:$ (2600-3000)$^*$ &   --   & 0.04184 & 0.0209	 &    --    &   --   &   --   & 0.01045  & 0.002688 & 0.00103  & 0.0033 \\ 
$m_{t\bar{t}}:$ (3000-3500)$^*$ &   --   & 0.01170 & 0.005851  &    --    &   --   &   --   & 0.002926 & 0.001271 & 0.000317 & 0.0015 \\ 
$m_{t\bar{t}}:$ (3500-4000)$^*$ &   --   & 0.00329 & 0.001643  &    --    &   --   &   --   & 0.000822 & 0.000674 & 0.000097 & 0.00077 \\ \hline
\end{tabular}%
}
\caption{We show the unfolded bin contents for the absolute parton-level differential cross-section measurement. The correlations are shown in Fig.~\ref{fig:corr_ttbar}.}
\end{table}

\begin{table}[h!]
\centering
\hspace*{-0.8 cm}\resizebox{1.15\textwidth}{!}{%
\begin{tabular}{|l|c|c|c|c|c|c|c|c|c|c|}
\hline
\multirow{3}{*}{$p p \rightarrow t \bar{t} $} &\multirow{3}{*}{ Measured (\%)}& \multirow{3}{*}{SM (\%)} & \multicolumn{4}{c|}{LHC Unc. (\%)}   &  \multicolumn{4}{c|}{HL-LHC Unc. (\%)}   \\ \cline{4-11}
& && \multirow{2}{*}{theo.} & \multicolumn{3}{c|}{exp.}  & \multirow{2}{*}{theo.} & \multicolumn{3}{c|}{exp.} \\ \cline{5-7} \cline{9-11}
& && & stat. & sys. + mod. & tot. & & stat. & sys. + mod. & tot.\\ \hline
$m_{t\bar{t}}:$ (0-500)       & 0.45  & 0.55  & 0.0770  & 0.309    & 0.340  & 0.460  & 0.0385 & 0.0667   & 0.170   & 0.183 \\ 
$m_{t\bar{t}}:$ (500-750)     & 0.51  & 0.72  & 0.1008  & 0.219    & 0.210  & 0.304  & 0.0504 & 0.0474   & 0.105   & 0.115 \\ 
$m_{t\bar{t}}:$ (750-1000)    & 1.00  & 0.79  & 0.1106  & 0.533    & 0.460  & 0.704  & 0.0553 & 0.1152   & 0.230   & 0.257 \\
$m_{t\bar{t}}:$ (1000-1500)   & 1.69  & 0.96  & 0.1344  & 0.776    & 0.270  & 0.822  & 0.0672 & 0.1677   & 0.135   & 0.215 \\ 
$m_{t\bar{t}}:$ (1500-2000)$^*$ &  --   & 1.01  & 0.1414  &    --    &   --   &   --   & 0.0707 & 0.5703   & 0.624   & 0.846 \\ 
$m_{t\bar{t}}:$ (2000-2500)$^*$ &  --   & 1.10  & 0.1540  &    --    &   --   &   --   & 0.0770 & 1.4750   & 0.683   & 1.625 \\ 
$m_{t\bar{t}}:$ (2500-3000)$^*$ &  --   & 1.85  & 0.2590  &    --    &   --   &   --   & 0.1295 & 5.4213   & 1.147   & 5.541 \\ \hline
\end{tabular}%
}
\caption{We show the values included for the charged asymmetry. The correlations are shown in Fig.~\ref{fig:corr_ttbar_AC}. }
\end{table}

\begin{table}[h!]
\centering
\hspace*{-0.8 cm}\resizebox{1.15\textwidth}{!}{%
\begin{tabular}{|l|c|c|c|c|c|c|c|c|c|c|c|c|}
\hline
\multirow{3}{*}{$p p \rightarrow t \bar{t} Z$} &\multirow{3}{*}{ Measured ($\text{fb}\cdot\text{GeV}^{-1}$)}& \multirow{3}{*}{SM ($\text{fb}\cdot\text{GeV}^{-1}$)} & \multicolumn{5}{c|}{LHC Unc. ($\text{fb}\cdot\text{GeV}^{-1}$)}   &  \multicolumn{5}{c|}{HL-LHC Unc. ($\text{fb}\cdot\text{GeV}^{-1}$)}   \\ \cline{4-13}
& && \multirow{2}{*}{theo.} & \multicolumn{4}{c|}{exp.}  & \multirow{2}{*}{theo.} & \multicolumn{4}{c|}{exp.} \\ \cline{5-8} \cline{10-13}
& && & stat. & sys. & mod. & tot. & & stat. & sys. & mod. & tot.\\ \hline
$p_T^Z:$ (0-40)      & 1.47  & 2.21 & 0.263  & 0.53 & 0.23  & 0.21  & 0.615 & 0.132 & 0.114 & 0.050 & 0.105 & 0.163  \\ 
$p_T^Z:$ (40-70)     & 4.32  & 4.59 & 0.543  & 0.94 & 0.60  & 0.51  & 1.223 & 0.272 & 0.203 & 0.130 & 0.253 & 0.349  \\ 
$p_T^Z:$ (70-110)    & 4.24  & 4.60 & 0.555  & 0.75 & 0.54  & 0.36  & 0.993 & 0.278 & 0.162 & 0.117 & 0.182 & 0.270 \\
$p_T^Z:$ (110-160)   & 4.4   & 3.45 & 0.429  & 0.55 & 0.43  & 0.39  & 0.800 & 0.215 & 0.118 & 0.093 & 0.197 & 0.248 \\ 
$p_T^Z:$ (160-220)   & 1.75  & 2.05 & 0.261  & 0.31 & 0.15  & 0.13  & 0.371 & 0.131 & 0.067 & 0.033 & 0.066 & 0.100 \\
$p_T^Z:$ (220-290)   & 0.58  & 1.03 & 0.130  & 0.16 & 0.047 & 0.034 & 0.174 & 0.065 & 0.035 & 0.010 & 0.017 & 0.041 \\
$p_T^Z:$ (290-400)   & 0.56  & 0.59 & 0.071  & 0.11 & 0.055 & 0.057 & 0.132 & 0.036 & 0.023 & 0.012 & 0.029 & 0.038 \\ \hline
\end{tabular}%
}
\caption{We show the unfolded bin contents for the absolute parton-level differential cross-section measurement.  The correlations are shown in Fig.~\ref{fig:corr_ttZ}. }
\end{table}

\begin{table}
\centering
\hspace*{-0.8 cm}\resizebox{1.15\textwidth}{!}{%
\begin{tabular}{|l|c|c|c|c|c|c|c|c|c|c|c|c|}
\hline
\multirow{3}{*}{$p p \rightarrow t \bar{t} \gamma$} &\multirow{3}{*}{ Measured ($\text{fb}\cdot\text{GeV}^{-1}$)}& \multirow{3}{*}{SM ($\text{fb}\cdot\text{GeV}^{-1}$)} & \multicolumn{5}{c|}{LHC Unc. ($\text{fb}\cdot\text{GeV}^{-1}$)}   &  \multicolumn{5}{c|}{HL-LHC Unc. ($\text{fb}\cdot\text{GeV}^{-1}$)}   \\ \cline{4-13}
& && \multirow{2}{*}{theo.} & \multicolumn{4}{c|}{exp.}  & \multirow{2}{*}{theo.} & \multicolumn{4}{c|}{exp.} \\ \cline{5-8} \cline{10-13}
& && & stat. & sys. & mod. & tot. & & stat. & sys. & mod. & tot.\\ \hline
$p_T^\gamma:$ (20-25)      & 1.782  & 1.670   & 0.066   & 0.116   & 0.168    & 0.108   & 0.231   & 0.033   & 0.025    & 0.036    & 0.054   & 0.070  \\ 
$p_T^\gamma:$ (25-30)      & 1.328  & 1.183   & 0.040   & 0.089   & 0.052    & 0.092   & 0.138   & 0.020   & 0.019    & 0.011    & 0.046   & 0.051  \\ 
$p_T^\gamma:$ (30-35)      & 0.966  & 0.8663  & 0.0302  & 0.072   & 0.026    & 0.060   & 0.097   & 0.0151  & 0.016    & 0.0056   & 0.030   & 0.0342  \\ 
$p_T^\gamma:$ (35-40)      & 0.705  & 0.6616  & 0.0205  & 0.058   & 0.015    & 0.042   & 0.0733  & 0.0103  & 0.0125   & 0.0032   & 0.021   & 0.0248  \\ 
$p_T^\gamma:$ (40-47)      & 0.474  & 0.4790  & 0.0160  & 0.04    & 0.0096   & 0.048   & 0.0629  & 0.0080  & 0.0086   & 0.0021   & 0.024   & 0.0254  \\ 
$p_T^\gamma:$ (47-55)      & 0.333  & 0.3464  & 0.0094  & 0.031   & 0.0067   & 0.017   & 0.0360  & 0.0047  & 0.0067   & 0.0014   & 0.0085  & 0.0109  \\ 
$p_T^\gamma:$ (55-70)      & 0.221  & 0.2188  & 0.0056  & 0.019   & 0.0038   & 0.0081  & 0.0210  & 0.0028  & 0.0041   & 0.00082  & 0.0041  & 0.0058  \\ 
$p_T^\gamma:$ (70-85)      & 0.122  & 0.1286  & 0.0031  & 0.014   & 0.0026   & 0.0069  & 0.0158  & 0.0016  & 0.0030   & 0.00056  & 0.0035  & 0.0046  \\ 
$p_T^\gamma:$ (85-132)     & 0.060  & 0.06037 & 0.0017  & 0.005   & 0.0014   & 0.0068  & 0.0086  & 0.00084 & 0.0011   & 0.00029  & 0.0034  & 0.0036  \\ 
$p_T^\gamma:$ (132-180)    & 0.020  & 0.02373 & 0.00077 & 0.003   & 0.00044  & 0.00080 & 0.00314 & 0.00039 & 0.00065  & 0.000095 & 0.00040 & 0.00077  \\ 
$p_T^\gamma:$ (180-300)    & 0.009  & 0.00790 & 0.00028 & 0.00045 & 0.000085 & 0.0014  & 0.00144 & 0.00014 & 0.000097 & 0.000018 & 0.00068 & 0.00069  \\  \hline
\end{tabular}
}
\caption{We show the unfolded bin contents for the absolute parton-level differential cross-section measurement.  The correlations are shown in Fig.~\ref{fig:corr_ttA}. }
\end{table}

\begin{table}
\centering
\hspace*{-0.8 cm}\resizebox{1.15\textwidth}{!}{%
\begin{tabular}{|l|c|c|c|c|c|c|c|c|c|c|c|c|}
\hline
\multirow{3}{*}{Process} &\multirow{3}{*}{ Measured (fb)}& \multirow{3}{*}{SM (fb)} & \multicolumn{5}{c|}{LHC Unc. (fb)}   &  \multicolumn{5}{c|}{HL-LHC Unc. (fb)}   \\ \cline{4-13}
& && \multirow{2}{*}{theo.} & \multicolumn{4}{c|}{exp.}  & \multirow{2}{*}{theo.} & \multicolumn{4}{c|}{exp.} \\ \cline{5-8} \cline{10-13}
& && & stat. & sys. & mod. & tot. & & stat. & sys. & mod. & tot.\\ \hline
$p p \rightarrow t \bar{t} H+ tHq$       & 640   & 664.3 & 41.7  & 90    & 40     & 70.7  & 121.2 & 20.9  & 19.4   & 8.6    & 35.4  & 41.3  \\ 
$p p \rightarrow t \bar{t} Z$            &  990  & 810.9 &  85.8 & 51.5  & 48.9   & 67.3  & 97.8  & 42.9  & 11.1   & 10.6   & 33.6  & 37.0  \\ 
$p p \rightarrow t \bar{t} \gamma$       & 39.6  & 38.5  & 1.76  & 0.8   & 1.25   & 2.16  & 2.62  & 0.88  & 0.17   & 0.27   & 1.08  & 1.13  \\
$p p \rightarrow tZq$                    & 111   & 102   & 3.5   & 13.0  & 6.1    & 6.2   & 15.7  & 1.75  & 2.09   & 0.98   & 3.1   & 3.87   \\ 
$p p \rightarrow t\gamma q$              & 115.7 & 81    & 4     & 17.1  & 21.1   & 21.1  & 34.4  & 2     & 1.9    & 2.3    & 10.6  & 11.0  \\
$p p \rightarrow t \bar{t} W+\text{EW}$  & 770   & 647.5 & 76.1  & 120   & 59.6   & 73.0  & 152.6 & 38.1  & 13.1   & 6.5    & 36.5  & 39.4  \\
$p p \rightarrow t\bar{b}$ (s-ch)        & 4900  & 5610  & 220   & 784   & 936    & 790   & 1454  & 110   & 35     & 42     & 395   & 399   \\ 
$p p \rightarrow tW$                     & 23100 & 22370 & 1570  & 1086  & 2000   & 2773  & 3587  & 785   & 49     & 89     & 1386  & 1390  \\ 
$p p \rightarrow tq$ (t-ch)              & 87700 & 84200 & 250   & 1140  & 3128   & 4766  & 5810  & 125   & 51     & 140    & 2383  & 2390  \\ 
$F_0$                                    & 0.693 & 0.687 & 0.005 & 0.009 & 0.006  & 0.009 & 0.014 & 0.003 & 0.0004 & 0.0003 & 0.004 & 0.004 \\ 
$F_L$                                    & 0.315 & 0.311 & 0.005 & 0.006 & 0.003  & 0.008 & 0.011 & 0.003 & 0.0003 & 0.0002 & 0.004 & 0.004 \\ \hline
\end{tabular}
}
\caption{The data shown is the inclusive cross-section written in fb for all the channels except for the $W$ Helicities ($F_0$ and $F_L$).}
\end{table}

\FloatBarrier

\begin{figure}\centering
\includegraphics[width=0.8\columnwidth]{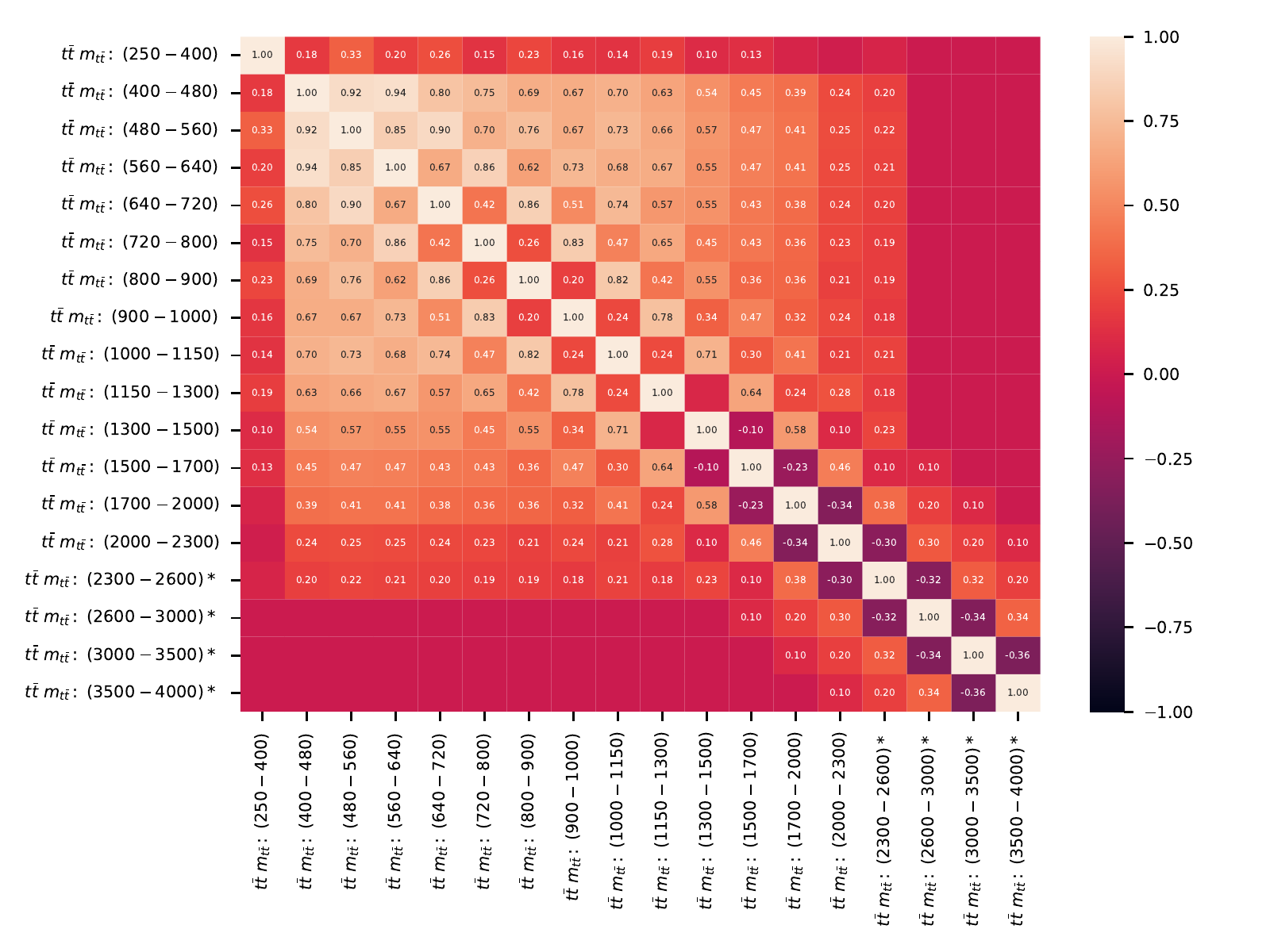}
\caption{\label{fig:corr_ttbar}%
Correlation matrix included for the differential cross section of $pp\to t \bar{t}$ extracted from \cite{CMS:2021vhb}. Cells are
filled if the correlation is higher than 10\% in absolute value. }
\end{figure}

\begin{figure}\centering
\includegraphics[width=0.8\columnwidth]{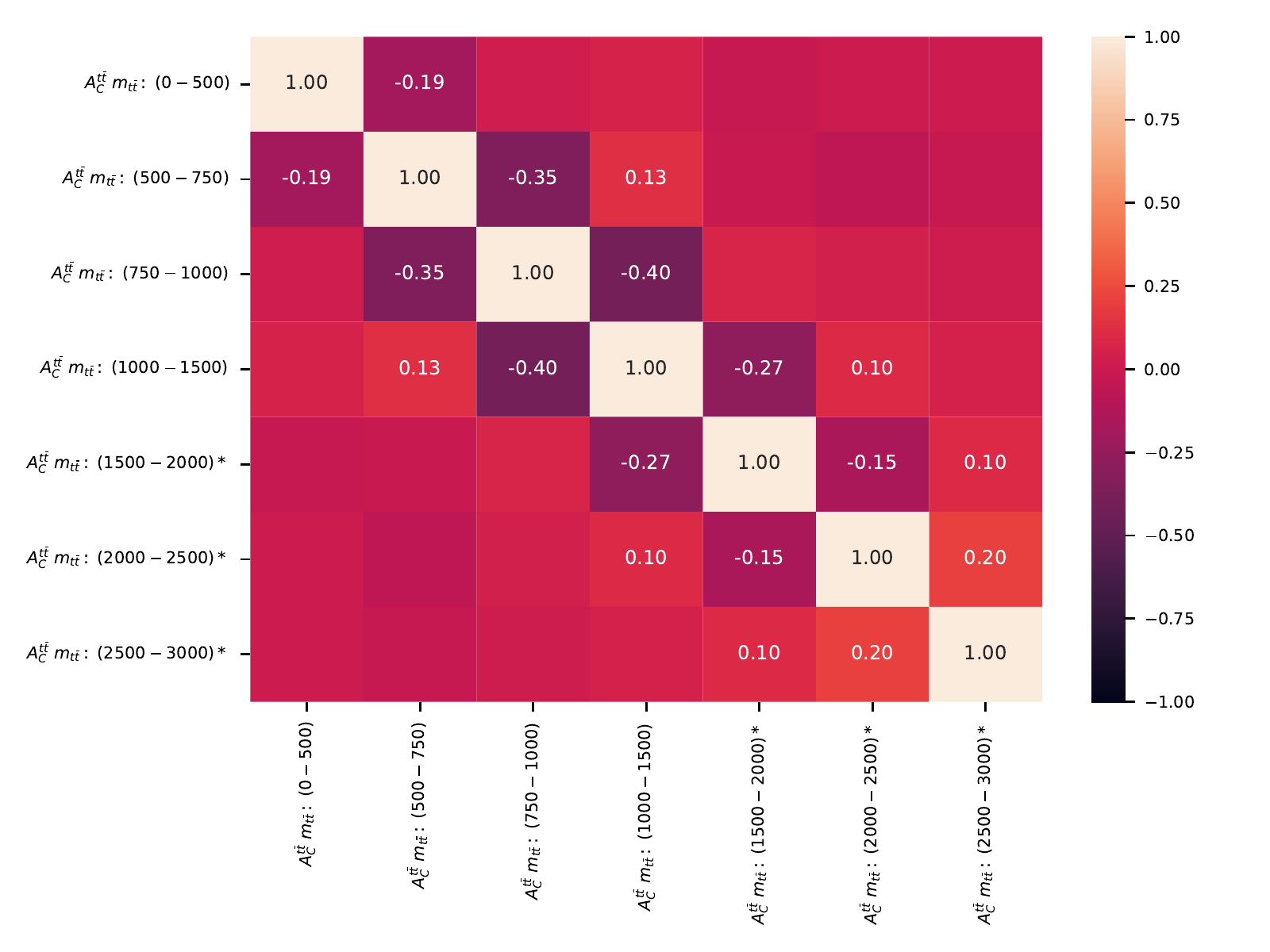}
\caption{\label{fig:corr_ttbar_AC}%
Correlation matrix included for the differential measurements of the charged asymmetry of $pp\to t \bar{t}$ extracted from \cite{ATLAS-CONF-2019-026}. Cells are filled if the correlation is higher than 10\% in absolute value.}
\end{figure}

\begin{figure}\centering
\includegraphics[width=0.8\columnwidth]{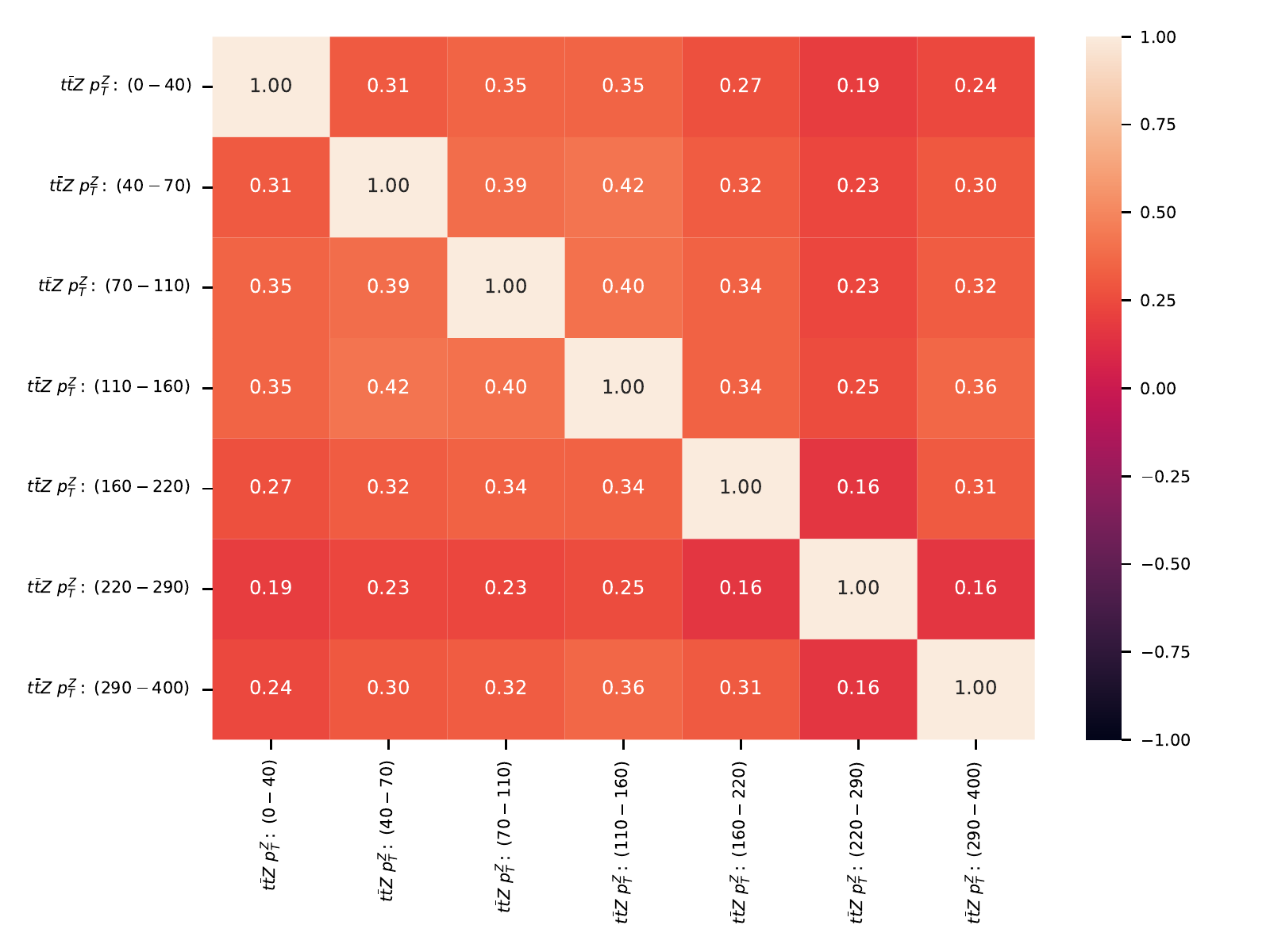}
\caption{\label{fig:corr_ttZ}%
Correlation matrix included for the differential measurements of the charged asymmetry of $pp\to t \bar{t} Z$ extracted from \cite{ATLAS:2020cxf}. Cells are filled if the correlation is higher than 10\% in absolute value. }
\end{figure}

\begin{figure}\centering
\includegraphics[width=0.8\columnwidth]{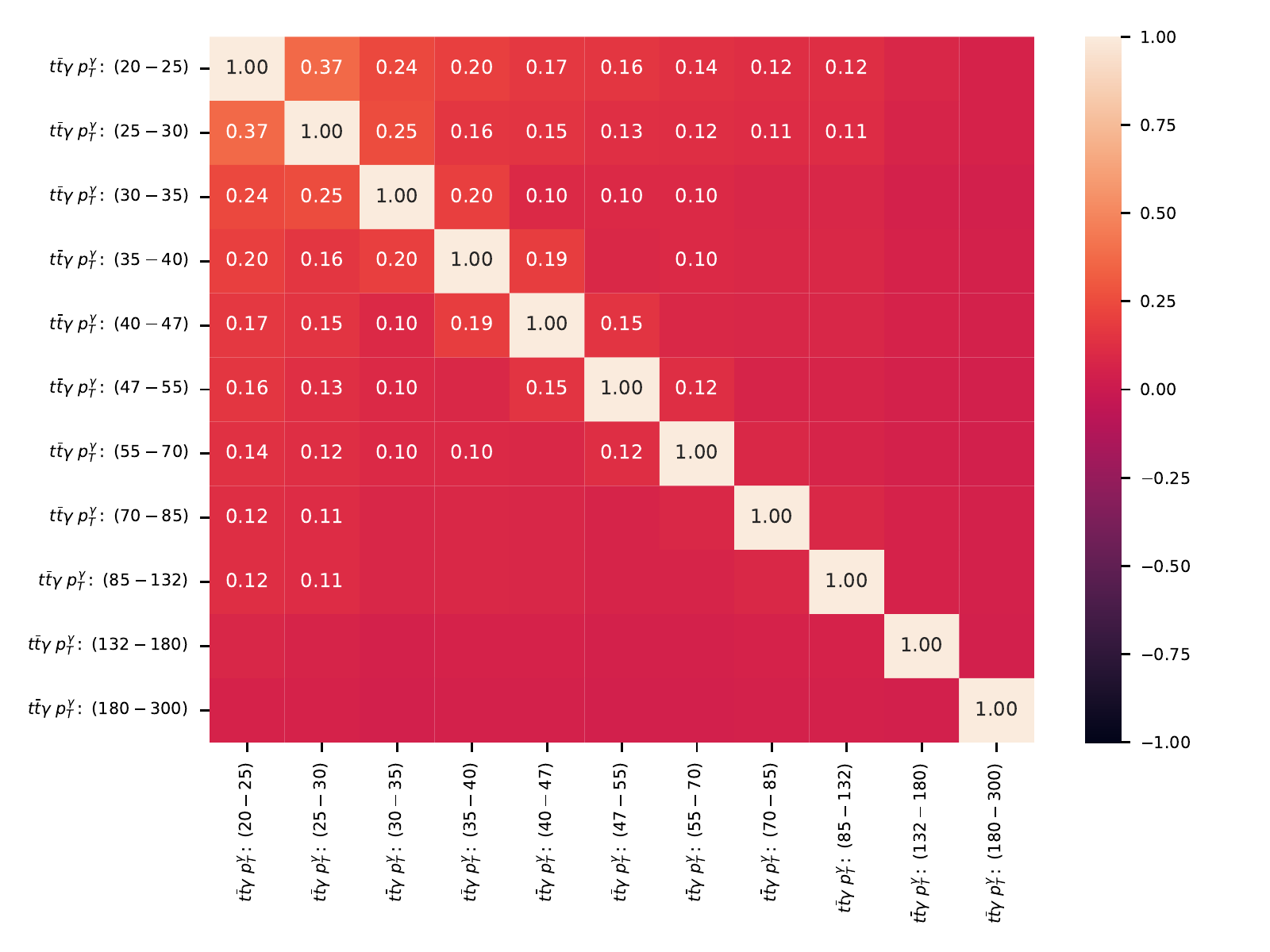}
\caption{\label{fig:corr_ttA}%
Correlation matrix included for the differential measurements of the charged asymmetry of $pp\to t \bar{t} \gamma$ extracted from \cite{Aad:2020axn}. Cells are filled if the correlation is higher than 10\% in absolute value. }
\end{figure}

\FloatBarrier

\subsection*{Future $e^+e^-$ colliders inputs: $e^+e^-\rightarrow b \bar{b}$}

\begin{table}[h!]
    \centering
\hspace*{-0.5 cm}\resizebox{1.1\textwidth}{!}{
    \begin{tabular}{c|c|c|c|c|c}
       Machine & Polarisation & Energy & Luminosity & Observable & Value\\\hline
        \multirow{12}{*}{ILC} & \multirow{6}{*}{{\scriptsize P($e^+$, $e^-$):$(-30\%,\,+80\%)$}}  &  \multirow{2}{*}{250 GeV} & \multirow{2}{*}{2 \iab{}} & $\sigma_b$ & $3182\pm 4.930  \text{ fb}$ \\\cline{5-6}
         &&&& $A_{\text{FB}}^b$ & $0.6267\pm 0.000738$ \\\cline{3-6}
         &  & \multirow{2}{*}{500 GeV} & \multirow{2}{*}{4 \iab{}} &  $\sigma_b$ & $ 693.5 \pm 1.188  \text{ fb}$ \\\cline{5-6}
           &&&& $A_{\text{FB}}^b$ & $0.6194\pm 0.001140$ \\\cline{3-6}
         &  & \multirow{2}{*}{1 TeV} & \multirow{2}{*}{8 \iab{}} & $\sigma_b$ & $168.10 \pm 0.311 \text{ fb}$ \\\cline{5-6}
          &&&& $A_{\text{FB}}^b$ & $0.617 \pm 0.001342$ \\\cline{2-6}\cline{2-6}
           & \multirow{6}{*}{{\scriptsize P($e^+$, $e^-$):$(+30\%,\,-80\%)$}}  &  \multirow{2}{*}{250 GeV} & \multirow{2}{*}{2 \iab{}} & $\sigma_b$ & $960.6\pm 1.774 \text{ fb}$ \\\cline{5-6}
         &&&& $A_{\text{FB}}^b$ & $0.426\pm 0.00201$ \\\cline{3-6}
         &  & \multirow{2}{*}{500 GeV} & \multirow{2}{*}{4 \iab{}} &  $\sigma_b$ & $ 203.9 \pm 0.442  \text{ fb}$ \\\cline{5-6}
           &&&& $A_{\text{FB}}^b$ & $0.493\pm 0.00268$ \\\cline{3-6}
         &  & \multirow{2}{*}{1 TeV} & \multirow{2}{*}{8 \iab{}} & $\sigma_b$ & $49.21 \pm 0.145$\\\cline{5-6}
          &&&& $A_{\text{FB}}^b$ & $0.507 \pm 0.00335$ \\\hline\hline

           \multirow{12}{*}{CLIC} & \multirow{6}{*}{{\scriptsize P($e^+$, $e^-$):$(0\%,\,+80\%)$}}  &  \multirow{2}{*}{380 GeV} & \multirow{2}{*}{1 \iab{}} & $\sigma_b$ & $964.6\pm2.03\text{ fb}$ \\\cline{5-6}
         &&&& $A_{\text{FB}}^b$ & $0.680 \pm 0.002$ \\\cline{3-6}
         &  & \multirow{2}{*}{1.5 TeV} & \multirow{2}{*}{2.5 \iab{}} &  $\sigma_b$ & $57.8\pm 0.311 \text{ fb}$ \\\cline{5-6}
           &&&& $A_{\text{FB}}^b$ & $0.674\pm 0.005$ \\\cline{3-6}
         &  & \multirow{2}{*}{3 TeV} & \multirow{2}{*}{5 \iab{}} & $\sigma_b$ & $14.41\pm 0.0923 \text{ fb} $\\\cline{5-6}
          &&&& $A_{\text{FB}}^b$ & $0.673 \pm 0.008$ \\\cline{2-6}
           & \multirow{6}{*}{{\scriptsize P($e^+$, $e^-$):$(0\%,\,-80\%)$}}  &  \multirow{2}{*}{380 GeV} & \multirow{2}{*}{1 \iab{}} & $\sigma_b$ & $329.8 \pm 1.01 \text{ fb}$\\\cline{5-6}
         &&&& $A_{\text{FB}}^b$ & $0.479\pm 0.004$\\\cline{3-6}
         &  & \multirow{2}{*}{1.5 TeV} & \multirow{2}{*}{2.5 \iab{}} &  $\sigma_b$ & $19.61\pm 0.153 \text{ fb}$ \\\cline{5-6}
           &&&& $A_{\text{FB}}^b$ & $0.520\pm 0.011$ \\\cline{3-6}
         &  & \multirow{2}{*}{3 TeV} & \multirow{2}{*}{5 \iab{}} & $\sigma_b$ & $4.88\pm 0.0505$\\\cline{5-6}
          &&&& $A_{\text{FB}}^b$ & $0.522\pm 0.015$ \\\hline\hline

          \multirow{6}{*}{FCC} & \multirow{8}{*}{Unpolaried}  & \multirow{2}{*}{Z-pole} & \multirow{2}{*}{150 \iab{}} & $\sigma_b$ & $8340800\pm 2449 \text{ fb}$ \\\cline{5-6}
        &&&& $A_{\text{FB}}^b$ & $0.23365	\pm 0.000130$\\\cline{3-6}
         &  & \multirow{2}{*}{240 GeV} & \multirow{2}{*}{5 \iab{}} & $\sigma_b$ & $1670\pm 2.51 \text{ fb}$ \\\cline{5-6}
         &&&& $A_{\text{FB}}^b$ & $0.584\pm 0.000679$\\\cline{3-6}
         &  & \multirow{2}{*}{365 GeV} & \multirow{2}{*}{1.5 \iab{}} & $\sigma_b$ & $647.2\pm 1.341 \text{ fb}$\\\cline{5-6}
          &&&& $A_{\text{FB}}^b$ & $0.591\pm 0.0016$ \\\hline

          \multirow{6}{*}{CEPC} & \multirow{8}{*}{Unpolaried}  & \multirow{2}{*}{Z-pole} & \multirow{2}{*}{57.5 \iab{}} & $\sigma_b$ & $8340800\pm 1947 \text{ fb}$ \\\cline{5-6}
        &&&& $A_{\text{FB}}^b$ & $0.23365	\pm 0.000130 $\\\cline{3-6}
         &  & \multirow{2}{*}{240 GeV} & \multirow{2}{*}{ 20 \iab{}} & $\sigma_b$ & $1670\pm 2.40 \text{ fb}$ \\\cline{5-6}
         &&&& $A_{\text{FB}}^b$ & $0.584\pm 0.000339 $\\\cline{3-6}
         &  & \multirow{2}{*}{365 GeV} & \multirow{2}{*}{1 \iab{}} & $\sigma_b$ & $647.2\pm 1.51 \text{ fb}$\\\cline{5-6}
          &&&& $A_{\text{FB}}^b$ & $0.591\pm 0.00200 $ \\\hline

    \end{tabular}
    }
    \caption{Inputs included to restrict the bottom quark sector}
    \label{tab:bigtable_epem_setup}
\end{table}

\FloatBarrier

\subsection*{Future $e^+e^-$ colliders inputs: $e^+e^-\rightarrow t \bar{t}$}

\begin{table}[h!]
\centering
\resizebox{\textwidth}{!}{
\begin{tabular}{c|c|c|cccc|}
\cline{2-7}
    \multirow{2}{*}{ ILC500 }  & \multirow{2}{*}{\makecell{Uncertainty\\Individual}} & \multirow{2}{*}{\makecell{Uncertainty\\Global}} & \multicolumn{4}{c|}{Correlation}\\\cline{4-7}
    & & & \multicolumn{1}{c}{$  C_{\varphi Q}^-$}  & \multicolumn{1}{c}{$C_{\varphi t}$}  & \multicolumn{1}{c}{$C_{t W}$} & \multicolumn{1}{c|}{ $C_{t Z}$} \\ \hline
    \multicolumn{1}{|c|}{$C_{\varphi Q}^-$}    &  0.032     & $0.11$         &   1.0000 & 0.8221 & 0.9412 & 0.9381      \\
    \multicolumn{1}{|c|}{$C_{\varphi t}$}      &  0.032   & $0.11$       &        0.8221 & 1.0000 & 0.9379 & 0.9409    \\
    \multicolumn{1}{|c|}{$C_{t W}$}            &  0.0007   & $0.014$      &      0.9411 & 0.9379 & 1.0000  & 0.9986    \\
    \multicolumn{1}{|c|}{$C_{t Z}$}           &   0.0008   & $0.016$         &    0.9381 & 0.9409 & 0.9986 & 1.0000         \\ \hline
\end{tabular}
}
\caption{Inputs included to restrict the top-quark sector in the ILC working at 500 GeV. These constrains come from an analysis with the optimal observables from Ref.~\cite{Durieux:2018tev}. The details of the ILC configuration can be found in Tab.~\ref{tab:epem_setup}.}
\end{table}

\begin{table}[h!]
\centering
\hspace*{-0.8 cm}\resizebox{1.15\textwidth}{!}{
\begin{tabular}{c|c|c|cccccccc|}
\cline{2-11}
    \multirow{2}{*}{ \makecell{ILC500+\\ILC1000} }   &  \multirow{2}{*}{\makecell{Uncertainty\\Individual} }  & \multirow{2}{*}{ \makecell{Uncertainty\\Global}} & \multicolumn{8}{c|}{Correlation}  \\\cline{4-11}
    &                         &                    & \multicolumn{1}{c}{$ C_{\varphi Q}^-$}  & \multicolumn{1}{c}{$ C_{\varphi t}$}  & \multicolumn{1}{c}{$C_{t W}$} & \multicolumn{1}{c}{ $C_{t Z}$}& \multicolumn{1}{c}{ $C_{l Q}^- $}& \multicolumn{1}{c}{ $C_{e Q}$}& \multicolumn{1}{c}{ $C_{l t} $}& \multicolumn{1}{c|}{ $C_{e t}$} \\ \hline
    \multicolumn{1}{|c|}{$C_{\varphi Q}^-$} &    0.0268    & 0.0759   &   1.000 &  0.306  & 0.734 &  0.720 &  -0.397 & 0.255   &  0.315 & -0.227   \\
    \multicolumn{1}{|c|}{$C_{\varphi t}$}   &    0.0268    & 0.0764   &   0.306 &  1.000  & 0.724 &  0.738 &  0.261  &  -0.317 & -0.354 & 0.326    \\
    \multicolumn{1}{|c|}{$C_{t W}$}         &    0.00057   & 0.0079   &   0.734 &  0.724  & 1.000 &  0.991 &  0.120  &  -0.160 &  0.178 & -0.084   \\
    \multicolumn{1}{|c|}{$C_{t Z}$}         &    0.00065   & 0.0090   &   0.720 &  0.738  & 0.991 &  1.000 &  0.093  &  -0.225 &  0.125 & -0.129   \\
    \multicolumn{1}{|c|}{$C_{l Q}^- $}      &    0.00024   & 0.00041  &  -0.397 & 0.261   & 0.120 & 0.093  & 1.000   &  -0.279 & -0.180 & 0.184    \\
    \multicolumn{1}{|c|}{$C_{e Q}$}         &    0.00034   & 0.00053  &   0.255 & -0.317  & -0.160 & -0.225 & -0.279 &  1.000  & 0.204  & -0.055   \\
    \multicolumn{1}{|c|}{$C_{l t} $}        &    0.00029   & 0.00048  &   0.315 & -0.354  & 0.178 &  0.125 &  -0.180 &  0.204  &  1.000 & -0.280   \\
    \multicolumn{1}{|c|}{$C_{e t}$}         &    0.00025   & 0.00040  &  -0.227 & 0.326   & -0.084 & -0.129 & 0.184  & -0.055  & -0.280 & 1.000   \\ \hline
\end{tabular}
}
\caption{Inputs included to restrict the top-quark sector in the ILC working at 500 GeV and 1000 GeV. These constrains come from an analysis with the optimal observables from Ref.~\cite{Durieux:2018tev}. The details of the ILC configuration can be found in Tab.~\ref{tab:epem_setup}.}
\end{table}

\begin{table}[h!]
\centering
\hspace*{-0.8 cm}\resizebox{1.15\textwidth}{!}{
\begin{tabular}{c|c|c|cccccccc|}
\cline{2-11}
    \multirow{2}{*}{ CLIC }   &  \multirow{2}{*}{\makecell{Uncertainty\\Individual} }  & \multirow{2}{*}{ \makecell{Uncertainty\\Global}} & \multicolumn{8}{c|}{Correlation}  \\\cline{4-11}
    &                         &                    & \multicolumn{1}{c}{$ C_{\varphi Q}^-$}  & \multicolumn{1}{c}{$ C_{\varphi t}$}  & \multicolumn{1}{c}{$C_{t W}$} & \multicolumn{1}{c}{ $C_{t Z}$}& \multicolumn{1}{c}{ $C_{l Q}^- $}& \multicolumn{1}{c}{ $C_{e Q}$}& \multicolumn{1}{c}{ $C_{l t} $}& \multicolumn{1}{c|}{ $C_{e t}$} \\ \hline
    \multicolumn{1}{|c|}{$C_{\varphi Q}^-$} &    0.065     & 0.127   & 1.000  &-0.097& 0.599 & 0.585 & -0.312 & 0.185 & 0.305 & -0.256   \\
    \multicolumn{1}{|c|}{$C_{\varphi t}$}   &    0.065     & 0.128   & -0.097 & 1.000 & 0.593 & 0.604 & 0.310 & -0.249 & -0.266 & 0.259    \\
    \multicolumn{1}{|c|}{$C_{t W}$}         &    0.00125   & 0.0114   & 0.599 & 0.593 & 1.000 & 0.992 & 0.110 & -0.118 & 0.136  & -0.082  \\
    \multicolumn{1}{|c|}{$C_{t Z}$}         &    0.00144   & 0.0130   & 0.585  & 0.604 & 0.992  & 1.000  & 0.098 & -0.148 & 0.112 & -0.104   \\
    \multicolumn{1}{|c|}{$C_{l Q}^- $}      &    0.00012   & 0.00015   & -0.312& 0.310 & 0.110  & 0.098  & 1.000 & -0.114 & -0.233& -0.026    \\
    \multicolumn{1}{|c|}{$C_{e Q}$}         &    0.00019   & 0.00021  &  0.185 & -0.249& -0.118 &-0.148 &-0.114 &1.000  &-0.066 &-0.162   \\
    \multicolumn{1}{|c|}{$C_{l t} $}        &    0.00015   & 0.00018  &  0.305 & -0.266 & 0.136 & 0.112 & -0.233& -0.066& 1.000 & -0.131   \\
    \multicolumn{1}{|c|}{$C_{e t}$}         &    0.00013   & 0.00015  &  -0.256 & 0.259 & -0.082& -0.104 &-0.026& -0.162& -0.131& 1.000   \\ \hline
\end{tabular}
}
\caption{Inputs included to restrict the top-quark sector in CLIC. These constrains come from an analysis with the optimal observables from Ref.~\cite{Durieux:2018tev}. The details of the CLIC configuration can be found in Tab.~\ref{tab:epem_setup}.}
\end{table}

\begin{table}[h!]
\centering
\hspace*{-0.8 cm}\resizebox{1.15\textwidth}{!}{
\begin{tabular}{c|c|c|cccccccc|}
\cline{2-11}
    \multirow{2}{*}{ FCC$-ee$ }   &  \multirow{2}{*}{\makecell{Uncertainty\\Individual} }  & \multirow{2}{*}{ \makecell{Uncertainty\\Global}} & \multicolumn{8}{c|}{Correlation}  \\\cline{4-11}
    &                         &                    & \multicolumn{1}{c}{$ C_{\varphi Q}^-$}  & \multicolumn{1}{c}{$ C_{\varphi t}$}  & \multicolumn{1}{c}{$C_{t W}$} & \multicolumn{1}{c}{ $C_{t Z}$}& \multicolumn{1}{c}{ $C_{l Q}^- $}& \multicolumn{1}{c}{ $C_{e Q}$}& \multicolumn{1}{c}{ $C_{l t} $}& \multicolumn{1}{c|}{ $C_{e t}$} \\ \hline
    \multicolumn{1}{|c|}{$C_{\varphi Q}^-$} &    0.143     & 10.91   & 1.000 & -0.678 & -0.045 & -0.213  & -0.996 & 0.979 & 0.703 & -0.595   \\
    \multicolumn{1}{|c|}{$C_{\varphi t}$}   &    0.147     & 10.91   & -0.678& 1.000  & -0.014 & -0.130  &  0.701 & -0.600 & -0.996& 0.979    \\
    \multicolumn{1}{|c|}{$C_{t W}$}         &    0.00145   & 0.0418  & -0.045& -0.014 & 1.000  & 0.610   & 0.051  & -0.116 & 0.017 & -0.090  \\
    \multicolumn{1}{|c|}{$C_{t Z}$}         &    0.00180   & 0.117   & -0.213& -0.130 &0.610 & 1.000  &0.149 & -0.393 &0.064 & -0.319   \\
    \multicolumn{1}{|c|}{$C_{l Q}^- $}      &    0.00293   & 0.381   & -0.996 & 0.701 & 0.051 & 0.149 & 1.000 & -0.959 &-0.721& 0.632    \\
    \multicolumn{1}{|c|}{$C_{e Q}$}         &    0.00454   & 0.365  &  0.979  &-0.600 &-0.116 &-0.393 &-0.959 &1.000  &0.639  &-0.486   \\
    \multicolumn{1}{|c|}{$C_{l t} $}        &    0.00300   & 0.384  &  0.703  &-0.996 &0.017  &0.064  &-0.721 &0.639  &1.000  &-0.958    \\
    \multicolumn{1}{|c|}{$C_{e t}$}         &    0.00427   & 0.358  &  -0.595 &0.979  &-0.090 &-0.319 &0.632 & -0.486 & -0.958& 1.000   \\ \hline
\end{tabular}
}
\caption{Inputs included to restrict the top-quark sector in the FCC$-ee$. These constrains come from an analysis with the optimal observables from Ref.~\cite{Durieux:2018tev}. The details of the FCC$-ee$ configuration can be found in Tab.~\ref{tab:epem_setup}.}
\end{table}

\begin{table}[h!]
\centering
\hspace*{-0.8 cm}\resizebox{1.15\textwidth}{!}{
\begin{tabular}{c|c|c|cccccccc|}
\cline{2-11}
    \multirow{2}{*}{ CEPC }   &  \multirow{2}{*}{\makecell{Uncertainty\\Individual} }  & \multirow{2}{*}{ \makecell{Uncertainty\\Global}} & \multicolumn{8}{c|}{Correlation}  \\\cline{4-11}
    &                         &                    & \multicolumn{1}{c}{$ C_{\varphi Q}^-$}  & \multicolumn{1}{c}{$ C_{\varphi t}$}  & \multicolumn{1}{c}{$C_{t W}$} & \multicolumn{1}{c}{ $C_{t Z}$}& \multicolumn{1}{c}{ $C_{l Q}^- $}& \multicolumn{1}{c}{ $C_{e Q}$}& \multicolumn{1}{c}{ $C_{l t} $}& \multicolumn{1}{c|}{ $C_{e t}$} \\ \hline
    \multicolumn{1}{|c|}{$C_{\varphi Q}^-$} &    0.172     & 11.55   & 1.000 & -0.683 &-0.044 &-0.230 &-0.995 &0.975  &0.714  &-0.584   \\
    \multicolumn{1}{|c|}{$C_{\varphi t}$}   &    0.178     & 11.50   & -0.683& 1.000  &-0.012 &-0.141 &0.712  &-0.591 &-0.995 &0.975    \\
    \multicolumn{1}{|c|}{$C_{t W}$}         &    0.00174   & 0.0501  & -0.044& -0.012 &1.000  &0.608  &0.052  &-0.118 &0.019  &-0.093  \\
    \multicolumn{1}{|c|}{$C_{t Z}$}         &    0.00217   & 0.137   & -0.230& -0.141 &0.608  &1.000  &0.163  &-0.423 &0.070  &-0.345   \\
    \multicolumn{1}{|c|}{$C_{l Q}^- $}      &    0.00353   & 0.403   & -0.995& 0.7120 & 0.052 & 0.163 & 1.000 & -0.951& -0.735& 0.628    \\
    \multicolumn{1}{|c|}{$C_{e Q}$}         &    0.00548   & 0.392  &  0.975 & -0.591 &-0.118 &-0.423 &-0.951 &1.000  &0.636  &-0.456   \\
    \multicolumn{1}{|c|}{$C_{l t} $}        &    0.00363   & 0.405  &  0.714 & -0.995 &0.019  &0.070  &-0.735 &0.636  &1.000  &-0.945    \\
    \multicolumn{1}{|c|}{$C_{e t}$}         &    0.00515   & 0.382  &  -0.584& 0.975  &-0.093 &-0.345 &0.628  &-0.456 &-0.950 &1.000   \\ \hline
\end{tabular}
}
\caption{Inputs included to restrict the top-quark sector in CEPC. These constrains come from an analysis with the optimal observables from Ref.~\cite{Durieux:2018tev}. The details of the CEPC configuration can be found in Tab.~\ref{tab:epem_setup}.}
\end{table}

\FloatBarrier

\section{Appendix: Correlation matrices}

\label{app:correlation_matrices}

In the following we show the correlation matrices obtained for the different scenarios that we have considered.

\begin{figure}\centering
\includegraphics[width=1.1\columnwidth]{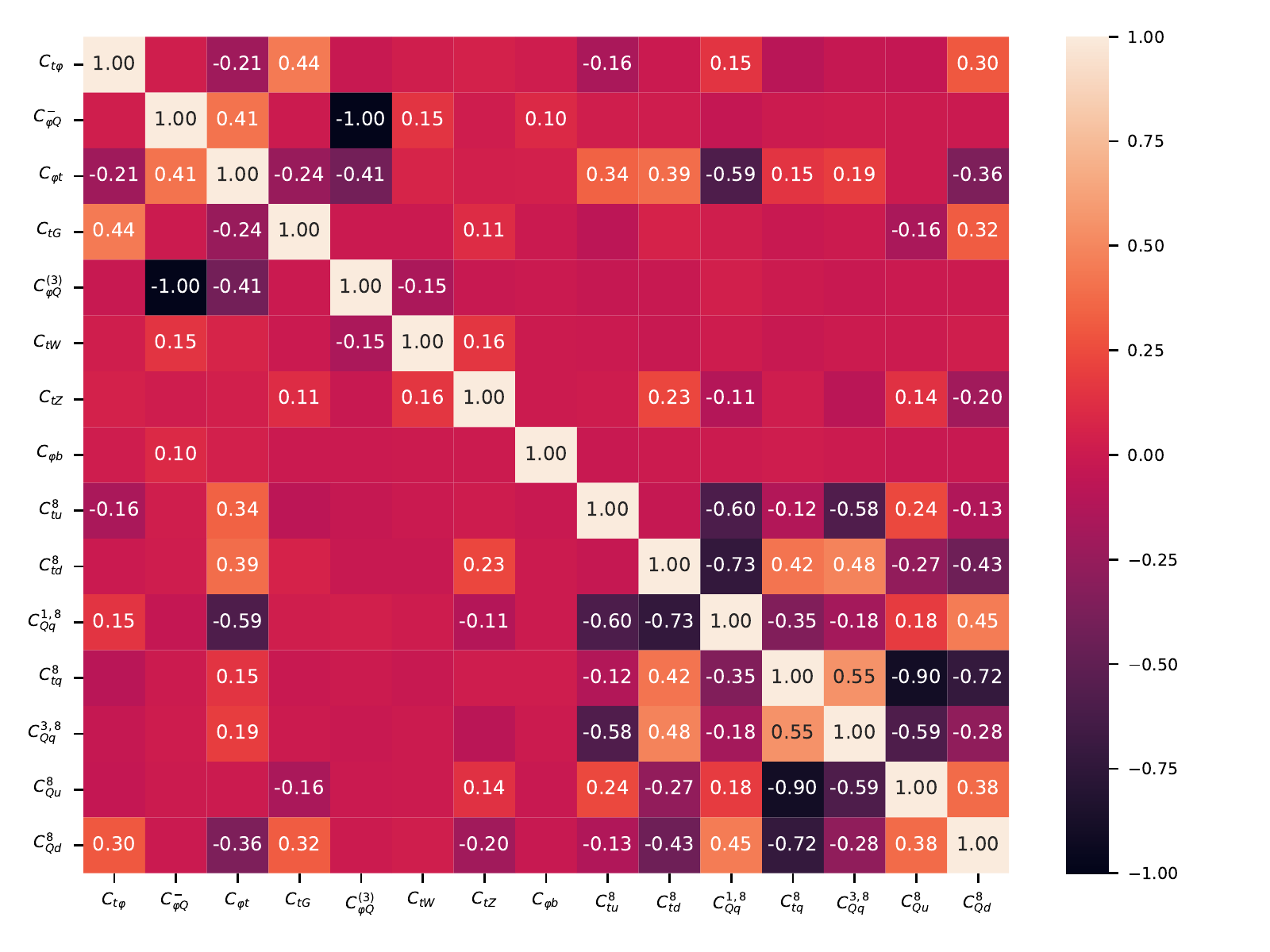}
\caption{\label{fig:corr_LHC_1}Correlation matrix obtained for the global fit including the data of the LHC, Tevatron and LEP. }
\end{figure}

\begin{figure}\centering
\includegraphics[width=1.1\columnwidth]{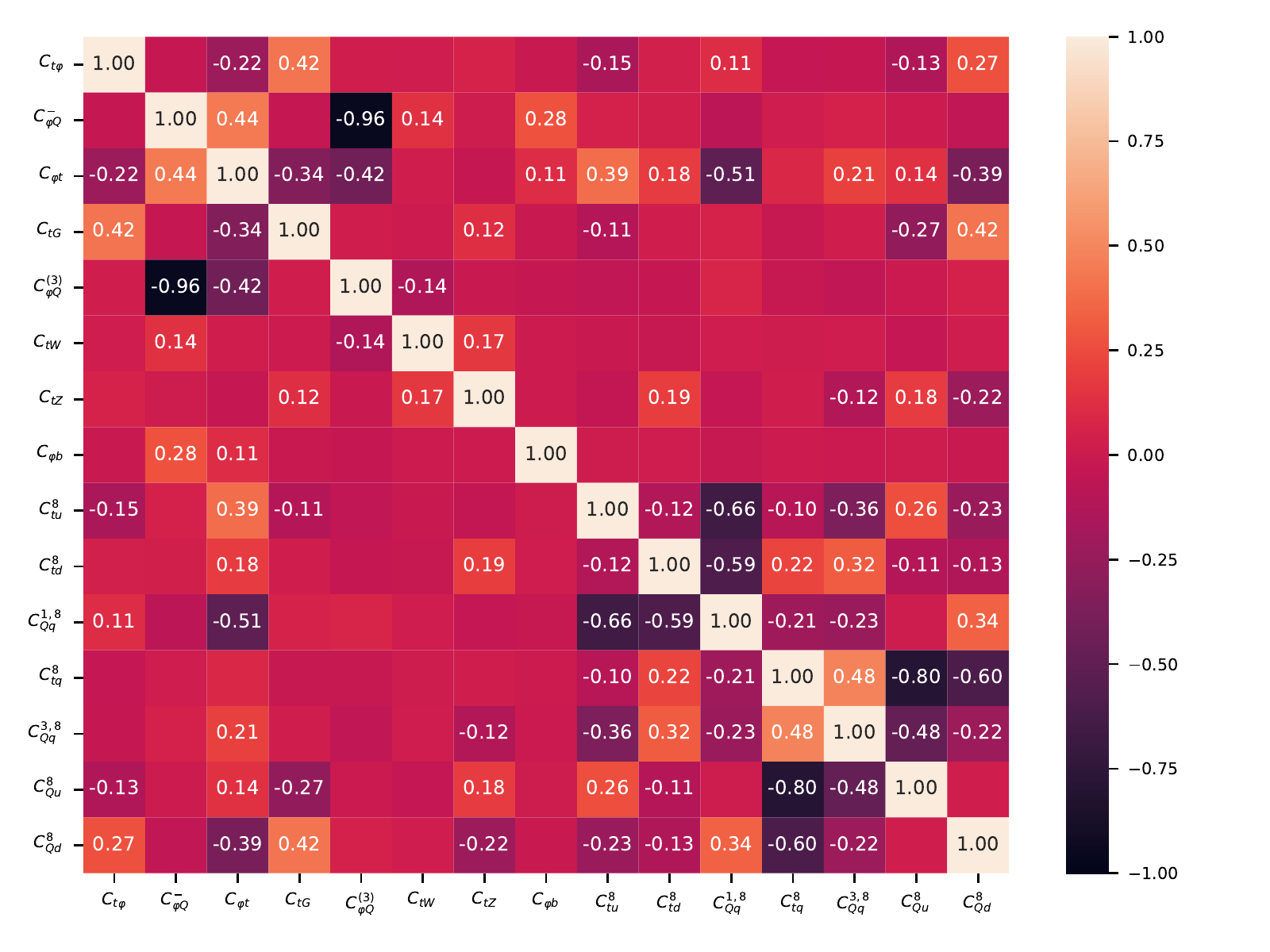}
\caption{\label{fig:corr_LHC_2}Correlation matrix obtained for the global fit including the data of the HL-LHC, Tevatron and LEP. }
\end{figure}

\begin{figure}\centering
\includegraphics[width=1.1\columnwidth]{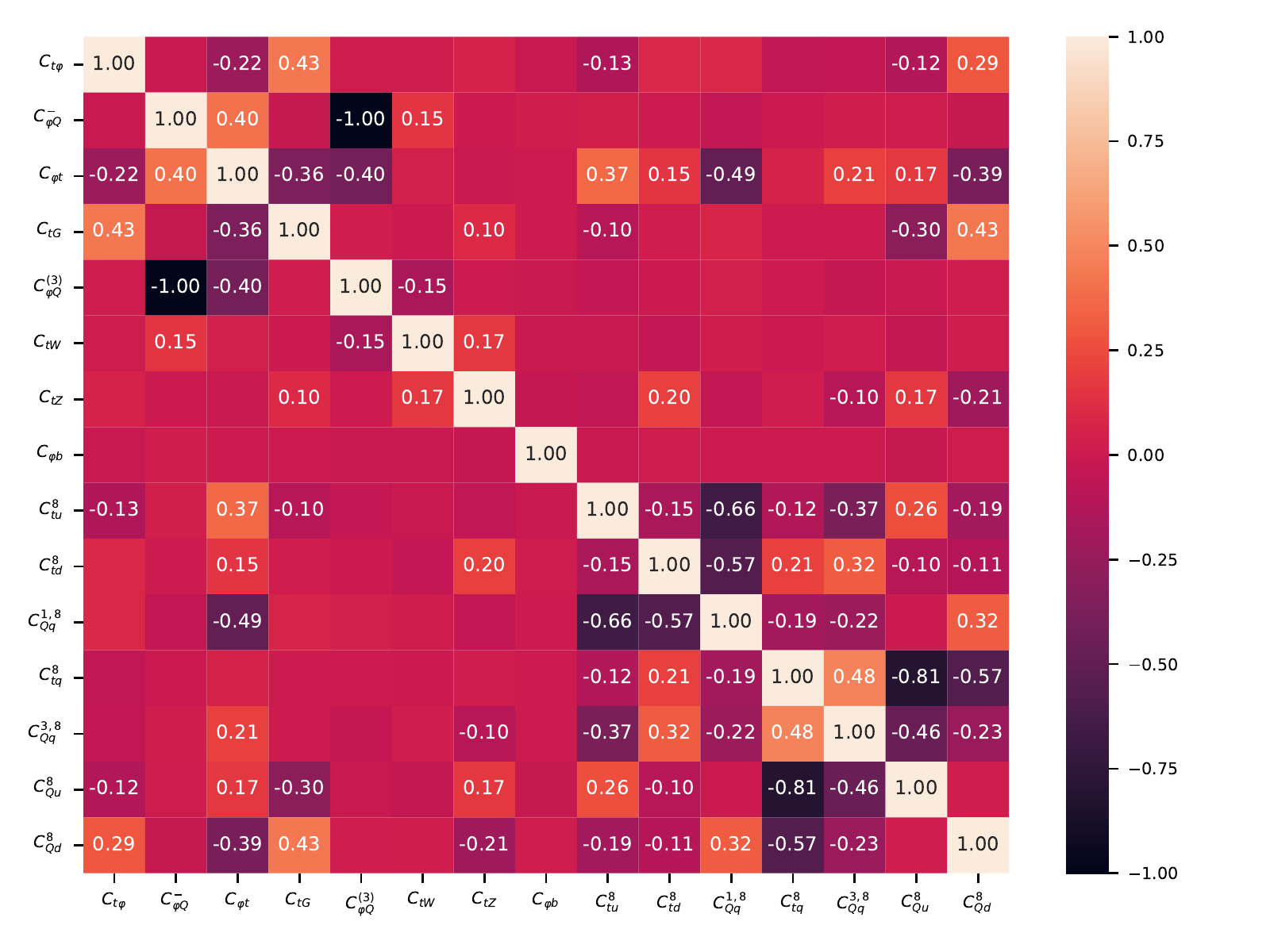}
\caption{\label{fig:corr_LHC_3}Correlation matrix obtained for the global fit including the data of the HL-LHC, Tevatron, LEP and ILC working at 250 GeV. }
\end{figure}

\begin{figure}\centering
\includegraphics[width=1.1\columnwidth]{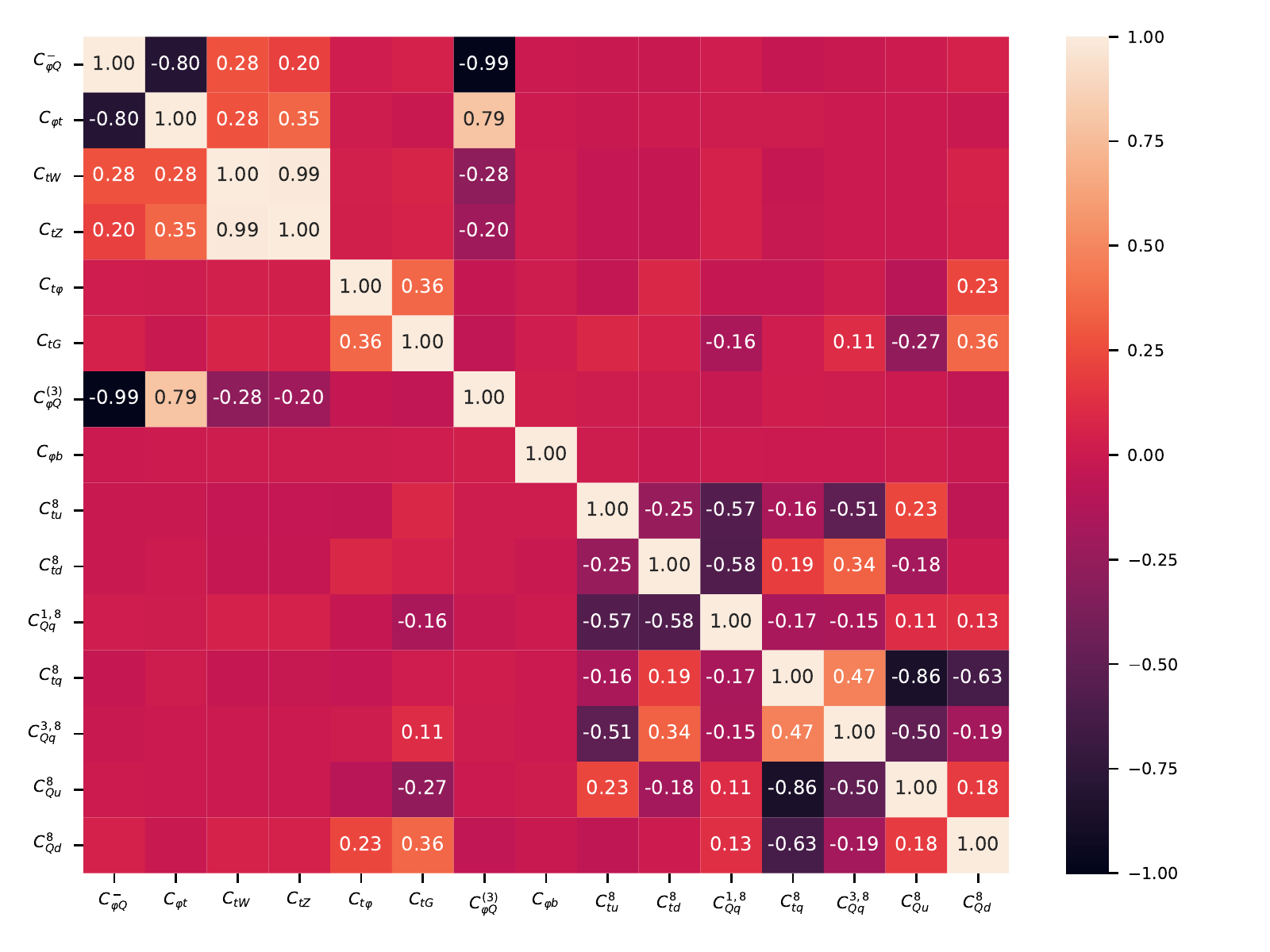}
\caption{\label{fig:corr_LHC_4}Correlation matrix obtained for the global fit including the data of the HL-LHC, Tevatron, LEP and ILC working at 250 GeV and 500 GeV. }
\end{figure}

\begin{figure}\centering
\includegraphics[width=1.1\columnwidth]{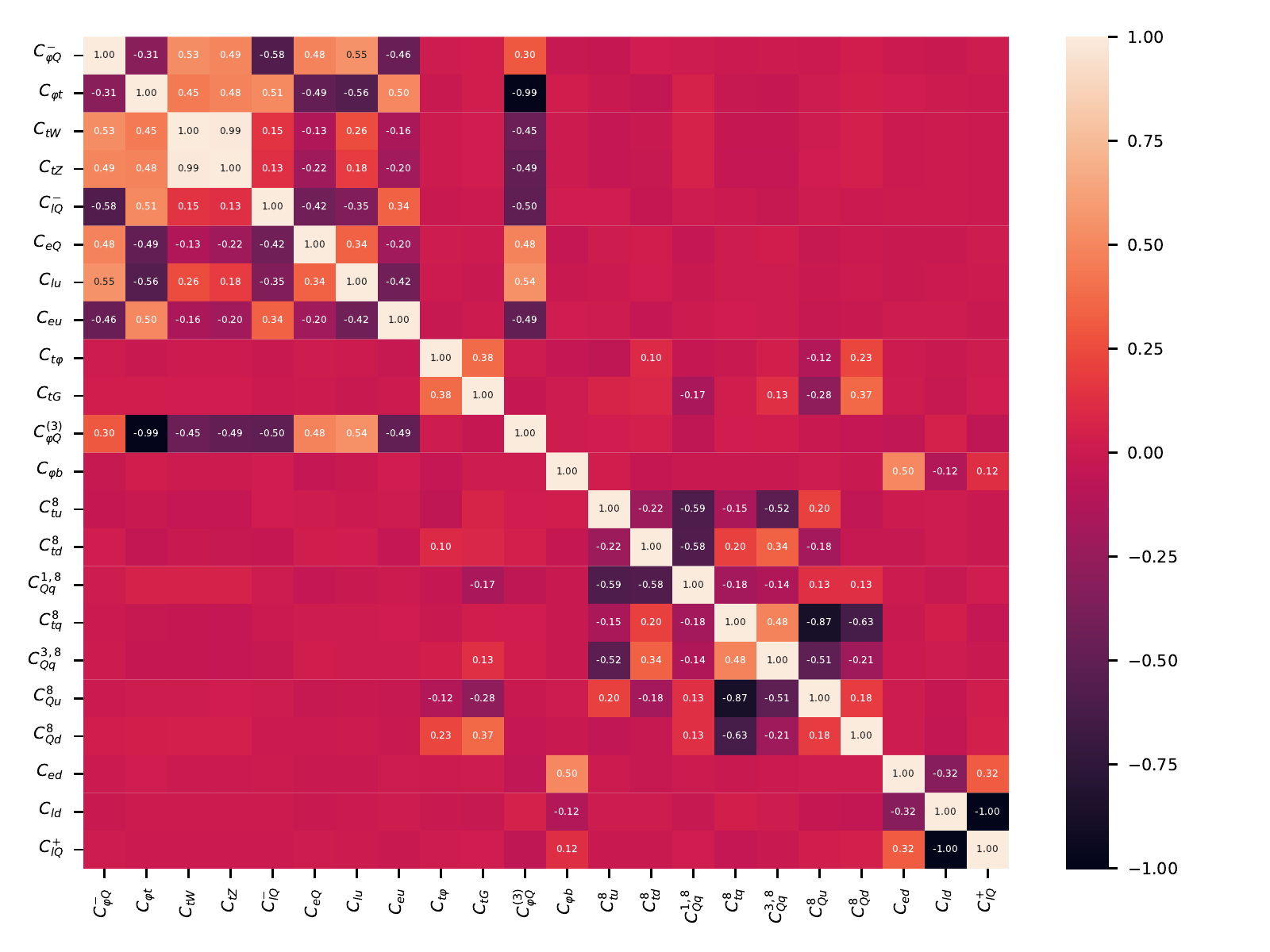}
\caption{\label{fig:corr_LHC_5}Correlation matrix obtained for the global fit including the data of the HL-LHC, Tevatron, LEP and ILC working at 250 GeV, 500 GeV and 1000 GeV. }
\end{figure}

\begin{figure}\centering
\includegraphics[width=1.1\columnwidth]{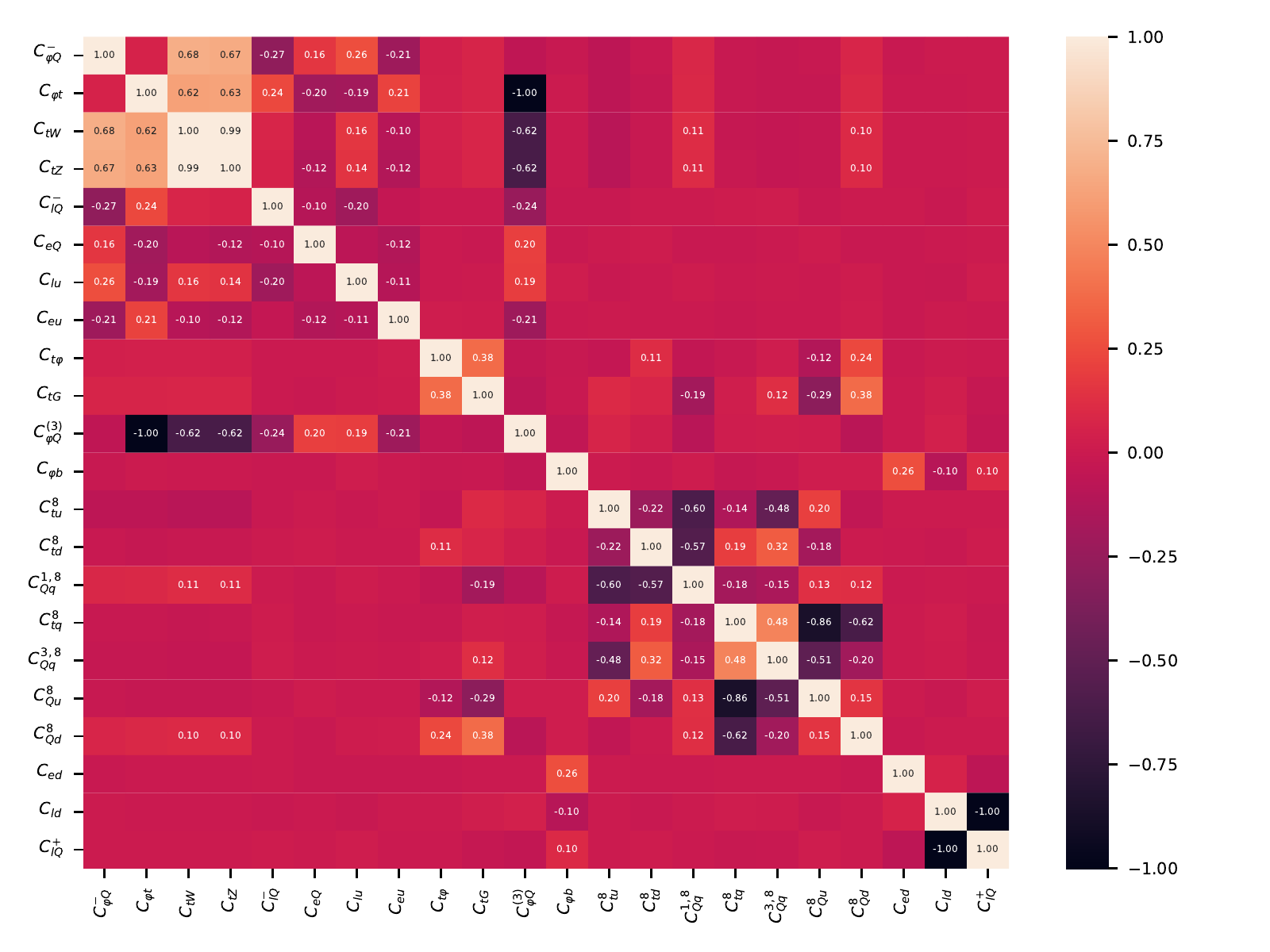}
\caption{\label{fig:corr_LHC_6}Correlation matrix obtained for the global fit including the data of the HL-LHC, Tevatron, LEP and the final stage of CLIC. }
\end{figure}

\begin{figure}\centering
\includegraphics[width=1.1\columnwidth]{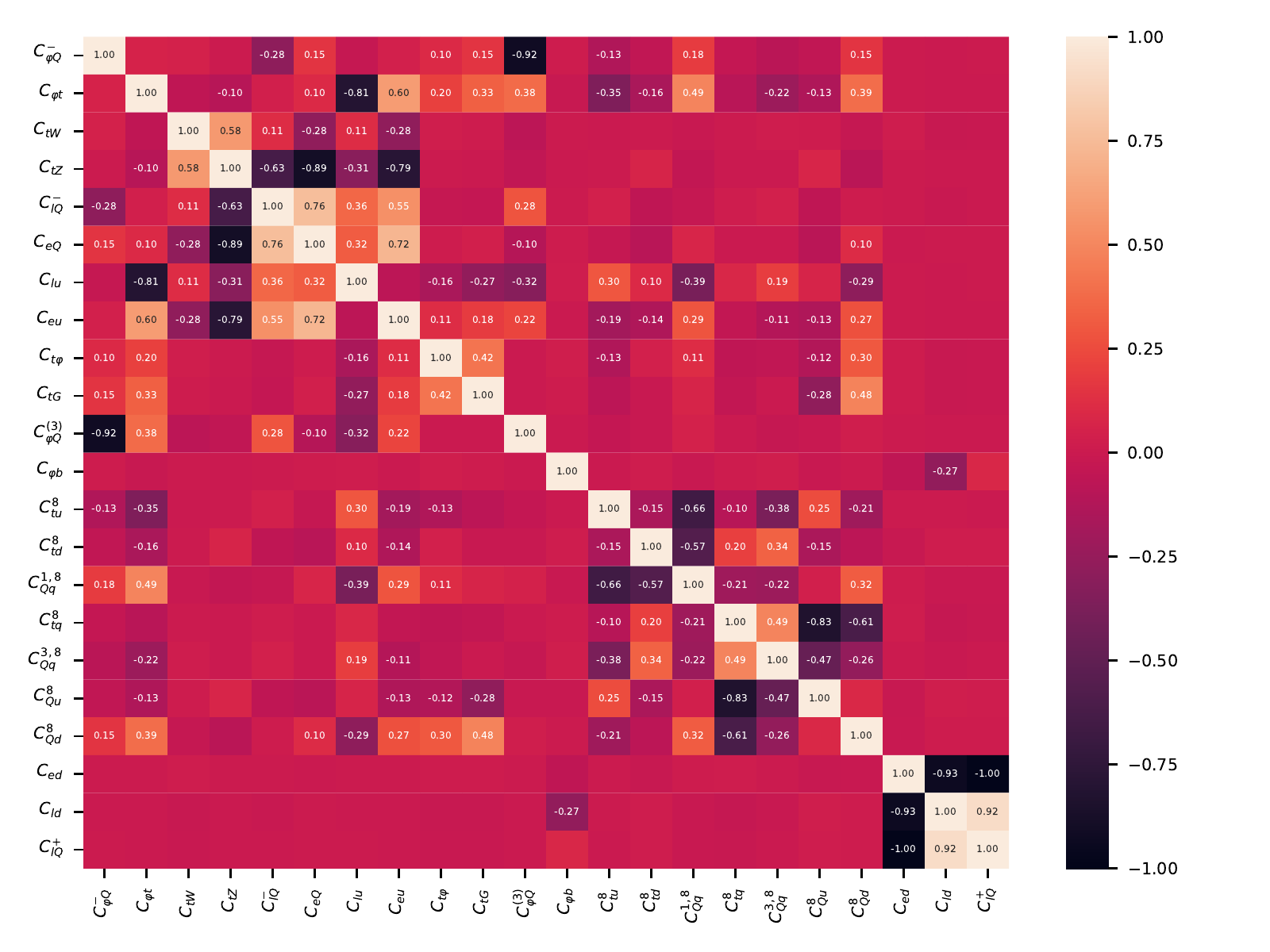}
\caption{\label{fig:corr_LHC_7}Correlation matrix obtained for the global fit including the data of the HL-LHC, Tevatron, LEP and the final stage of CEPC. }
\end{figure}

\begin{figure}\centering
\includegraphics[width=1.1\columnwidth]{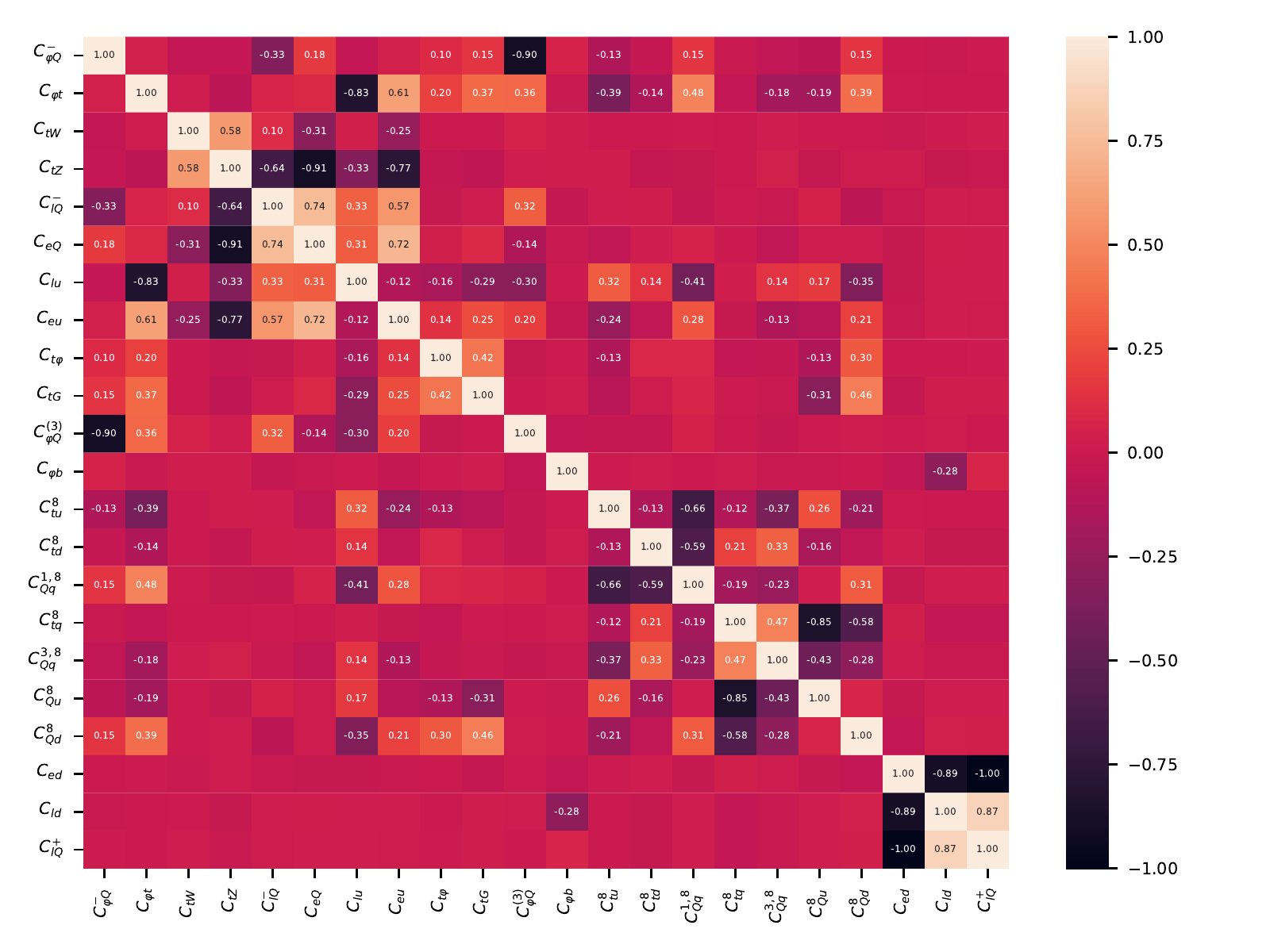}
\caption{\label{fig:corr_LHC_8}Correlation matrix obtained for the global fit including the data of the HL-LHC, Tevatron, LEP and the final stage of FCC. }
\end{figure}

\FloatBarrier

\newpage

\section{Appendix: Binning for the differential measurements} 
\label{app:differential}

The binning of the cross section measurement is based on the CMS measurement of Ref.~\cite{CMS:2021vhb}, combining analyses targetting the resolved and boosted topologies. This analysis is available on HEPDATA under: \url{https://www.hepdata.net/record/ins1901295}

The binning of the charge asymmetry measurement is based on the ATLAS run 2 analysis reported in Ref.~\cite{ATLAS-CONF-2019-026}.

In both cases, the $m_\ttbar$ range is extended with several further bins to take advantage of the greater reach of the full HL-LHC programme and the energy-growth of the sensitivity of some operators. The final binning is given in Table~\ref{tab:differential_measurements_binning}. Bins that have changed with respect to Refs.~\cite{CMS:2021vhb,ATLAS-CONF-2019-026}, and additional bins, are indicated with an asterisk (*).

\begin{table}[h!]
    \centering
    \resizebox{\textwidth}{!}{
    \begin{tabular}{|c|cccccc|}
    \hline
    observable     &  \multicolumn{6}{c|}{binning} \\  \hline
     $\sigma$ vs. $m_\ttbar$ [GeV] & bin 1 & bin 2 & bin 3 & bin 4 & bin 5 & bin 6 \\
                   & 250-400 & 400-480 & 480-560   & 560-640 & 640-720 & 720-800 \\
                   & bin 7 & bin 8 & bin 9 & bin 10 & bin 11 & bin 12  \\  
                   & 800-900 & 900-1000 & 1000-1150 & 1150-1300 & 1300-1500 & 1500-1700 \\
                   & bin 13 & bin 14 & bin 15 & bin 16 & bin 17 & bin 18  \\  
                   & 1700-2000 & 2000-2300 & 2300-2600$^*$ & 2600-3000$^*$ & 3000-3500$^*$ & 3500-4000$^*$ \\
     $A_C$ vs. $m_\ttbar$ [GeV]   & bin 1 & bin 2 & bin 3 & bin 4 & bin 5 & bin 6  \\
               & 500-750 & 750-1000 & 1000-1500 & 1500-2000$^*$ & 2000-2500$^*$ & 2500-3000$^*$ \\ \hline 
    \end{tabular}
    }
    \caption{The binning for the cross section and charge asymmetry differential measurements in $pp \rightarrow \ttbar$. Bins that differ from those used in the run 2 analyses are indicated with an asterisk (*). }
    \label{tab:differential_measurements_binning}
\end{table}

\newpage

\bibliographystyle{apsrev4-1_title}
\bibliography{top.bib}

\end{document}